\documentclass[preprint]{aastex63}

\received{}
\revised{}
\accepted{}
\submitjournal{PSJ}

\shorttitle{Titan tides}
\shortauthors{Idini \& Nimmo}
\graphicspath{{./}{figures/}}

\usepackage{bm}

\begin{document}

\title{Resonant stratification in Titan's global ocean}

\correspondingauthor{Benjamin Idini}
\email{bidini@ucsc.edu}

\author[0000-0002-2697-3893]{Benjamin Idini}
\affiliation{Department of Earth and Planetary Sciences, University of California Santa Cruz \\
1156 High St \\
Santa Cruz, CA 95064, USA}

\author[0000-0001-9432-7159]{Francis Nimmo}
\affiliation{Department of Earth and Planetary Sciences, University of California Santa Cruz \\
1156 High St \\
Santa Cruz, CA 95064, USA}



\begin{abstract}
    
Titan's ice shell floats on top of a global ocean revealed by the large tidal Love number $k_2 = 0.616\pm0.067$ registered by Cassini. The Cassini observation exceeds the predicted $k_2$ by one order of magnitude in the absence of an ocean, and is 3-$\sigma$ away from the predicted $k_2$ if the ocean is pure water resting on top of a rigid ocean floor. Previous studies demonstrate that an ocean heavily enriched in salts (\added{salinity} $S\gtrsim200$ g/kg) can explain the 3-$\sigma$ signal in $k_2$. Here we revisit previous interpretations of Titan's large $k_2$ using simple physical arguments and propose a new interpretation based on the dynamic tidal response of a stably stratified ocean in resonance with eccentricity tides raised by Saturn. Our models include inertial effects from a full consideration of the Coriolis force and the radial stratification of the ocean, typically neglected or approximated elsewhere. The stratification of the ocean emerges from a salinity profile where salt concentration linearly increases with depth. We find multiple salinity profiles that lead to the $k_2$ required by Cassini. In contrast with previous interpretations that neglect stratification, resonant stratification reduces the bulk salinity required by observations by an order of magnitude, reaching a salinity for Titan's ocean that is compatible with that of Earth's oceans and close to Enceladus' plumes. Consequently, no special process is required to enrich Titan's ocean to a high salinity as previously suggested. 

\end{abstract}




\section{Introduction} \label{sec:intro}

Recent decades of space exploration have revealed a solar system populated with internally heated icy worlds where large reservoirs of liquid water accumulate in subsurface global oceans \citep{nimmo2016ocean}. These worlds signal a possibility for life beyond Earth in a location that is accessible to future in-situ space exploration. A global ocean plays a fundamental role in determining the potential habitability of these icy worlds because water is required for life as we know it. Beyond detection, however, these global oceans remain poorly understood. Ocean thickness is typically known within broad limits ranging in the tens of percent of the icy world radius \citep{sohl2003interior,grindrod2008long}, preventing an accurate assessment of the satellite's liquid water inventory and thermal history. On Earth, ocean dynamics  modulate the distribution of nutrients and energy sources required by life \citep[e.g.]{Uchida:2020}, but on icy worlds the type of convection or lack thereof remains poorly known \citep{jansen2023energetic}. Here we argue in favor of the stratification of Titan's ocean based on a new interpretation of Cassini gravity measurements where internal gravity waves in Titan's ocean become resonantly excited by tides raised by Saturn (see also \cite{luan2019titan}). We hereafter refer to this proposed scenario as resonant stratification. 

Titan is the second largest solar system icy satellite and the best characterized from the perspective of gravity measurements (i.e., moment of inertia, $J_{2}$, $C_{22}$, and $k_2$ \citep{durante2019titan}), offering us a unique opportunity to reveal the hidden interior structure and dynamics of the global ocean within. The Cassini spacecraft unambiguously signaled the existence of a global ocean from the observed tidal response registered in the Love number $k_2 = 0.616\pm0.067$ \citep{durante2019titan}. The observation is an order of magnitude larger than the predicted $k_2\approx0.03$ when the ocean is absent \citep{rappaport2008can}. Ignoring the ice shell, a global ocean of pure liquid water produces a $k_2\approx 0.468$ independent of ocean thickness if the high-pressure ice and silicates beneath the ocean behave approximately rigidly and the total mass of the satellite is conserved (Section~\ref{sec:k2core}). The presence of an overlying elastic ice shell further restricts the motion of the ocean beneath. Estimates of Titan's ice shell thickness yield $d\sim100$~km \citep{sohl2003interior,nimmo2010shell,luan2019titan}; an ice shell this thick reduces tides down to $k_2\approx 0.42$ (Section~\ref{sec:shell}), which is roughly $3\sigma$ away from the Cassini observation. A thinner ice shell provides a $k_2$ closer to the observation, but then the heat conducted across the ice shell exceeds the interior heat production expected from radiogenic and tidal heating \citep{sohl2003interior,luan2019titan}, leading to thickening of the ice shell by freezing over time. A thinner shell is also more difficult to reconcile with the observed topography \citep{nimmo2010shell}.  The resonant stratification presented here can self-consistently explain the large $k_2$ observed by Cassini by introducing a positive dynamical fractional correction to the non-resonant hydrostatic $k_2\approx0.42$.

\begin{figure}[ht]
    \centering
    \plotone{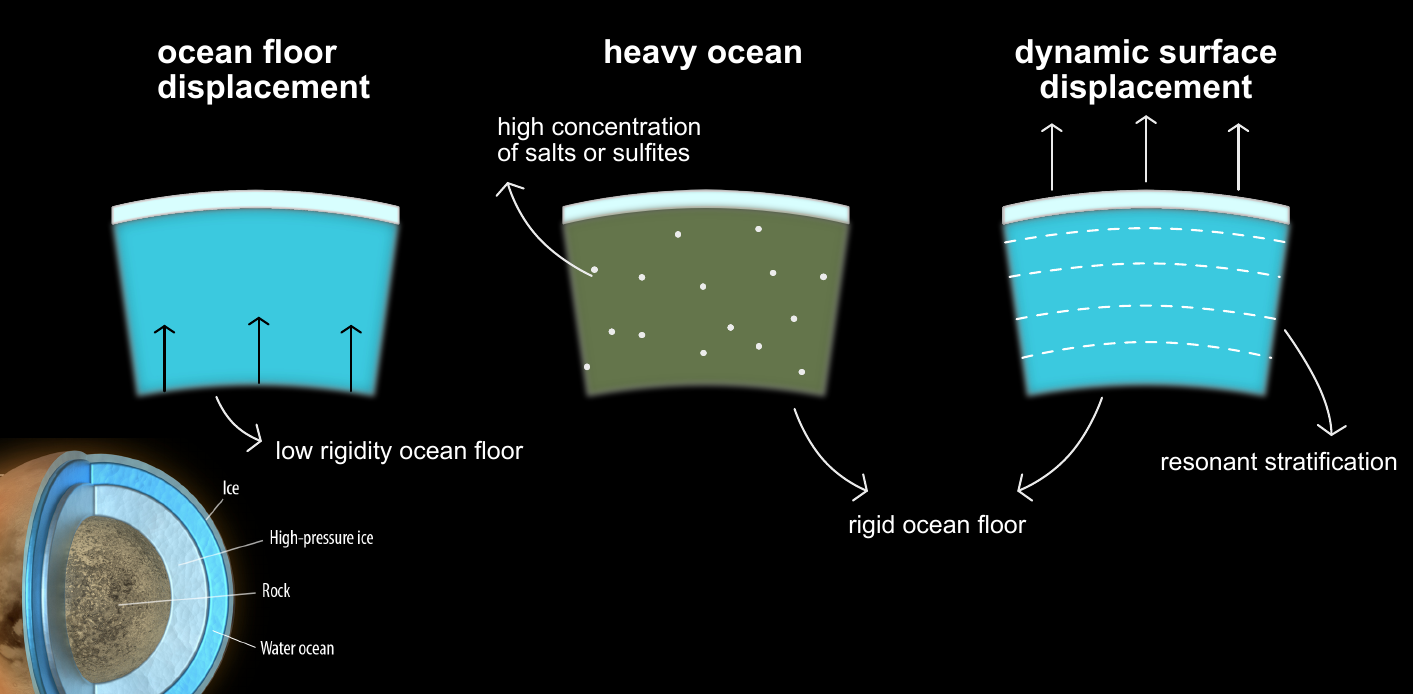}
    \caption{Proposed explanations to Titan's large $k_2$ as observed by Cassini. The schematic of Titan's interior model has been extracted from \cite{de2019tidal}. \label{fig:scenarios}}
\end{figure}

Resonant stratification enhances $k_2$ by dynamic amplification of the vertical displacement of Titan's surface (Fig.~\ref{fig:scenarios}). Ocean waves can produce significant dynamic surface displacement when resonantly excited by Saturn's eccentricity tides, namely when Titan's orbital frequency ($\omega_s=4.561\times10^{-6}$~s$^{-1}$) is close to a match with a normal mode emerging from an aggregation of Titan's ocean waves. Amplification results from the constructive addition of ocean waves over many cycles of resonant excitation, until balanced by dissipation. The effect produces dynamic gravity that can be registered by the tracking system of a nearby passing spacecraft (i.e., Cassini). A vertical gradient in ocean salinity promotes ocean stratification and the emergence of internal gravity waves. These waves are restored by buoyancy forces and organize in a spectrum of low-frequency normal modes that asymptotically approach zero frequency from an upper bound frequency $\omega^2\lesssim N^2$, where $N^2$ is the Brunt-Vaisala frequency that typically describes the strength of ocean stratification (Section~\ref{sec:dyn_tides} and Appendix~\ref{app:modes}). A typical value $N^2\gtrsim10^{-8}$~s$^{-2}\gtrsim\omega_s^2$ suggests a spectrum of internal gravity waves capable of resonance with eccentricity tides, and roughly corresponds to a modest vertical salinity gradient where salt concentration increases by ${\gtrsim} +1$ g/kg every ${\sim}100$ km of ocean thickness (the salinity of Earth's oceans is roughly $35$~g/kg). Crucially, our mechanism works for a range of mean ocean salinity values, including  Earth's and that inferred for Enceladus \citep{postberg2009sodium}. 

When the ocean is fully convective (i.e., unstratified) and homogeneously mixed, ocean waves require an unphysically thin ocean ($H <1$ km) to resonate with eccentricity tides  \citep{matsuyama2018ocean}. An ocean this thin is not compatible with the thermal history of Titan as expected from radiogenic and tidal heating \citep{tobie2006episodic,grindrod2008long}. 
Only the high-frequency tides from moon-to-moon interactions can resonantly excite a fully mixed ocean of realistic thickness ($H\sim100$~km)\citep{hay2020powering}, but those signals are relatively small and thus beyond Cassini's detection threshold. 

Alternative scenarios can enhance $k_2$ to the value required by Cassini (Fig.~\ref{fig:scenarios}), but require either high salinity of the ocean or low rigidity of the solid Titan. Previous calculations \citep{rappaport2008can,waite2017cassini} show that a heavy ocean produces the required extra gravity when its density is increased by a high concentration of dissolved salts (\added{salinity} $S \sim200$ g/kg). The required concentration of salts is an order of magnitude higher than that in Earth's oceans, Enceladus' plumes, and that predicted from water-rock interactions on Titan's ocean floor \citep{postberg2009sodium,leitner2019modeling}. On the other hand, a low rigidity or low viscosity ocean floor can allow large vertical displacements of the ocean floor that produce an extra gravity signal that constructively adds to the gravity from tides at the surface \citep{durante2019titan}. In practice, reasonable estimates of the rigidity and viscosity for silicates and high-pressure ice at tidal timescales produce a negligible ocean floor displacement. For example, the contribution to $k_2$ from the elastic icy ocean floor displacement can be estimated to be negligible following $k_{2,\textnormal{ice}}/k_2\times(\rho_{\textnormal{hp-ice}}-\rho)/\rho\sim 2\%$, where $k_{2,\textnormal{ice}}/k_2$ is the ratio between the Love number when the ocean is excluded/included, respectively, $\rho_{\textnormal{hp-ice}}$ is the density of high-pressure ice, and $\rho$ the ocean density. \added{The previous estimate assumes that the tidal response of the ocean floor can be crudely approximated by the tidal response of an oceanless Titan model after introducing a correction for the relatively higher surface density of high-pressure ice.} A viscous icy ocean floor can introduce a non-negligible component to $k_2$ if the viscosity of high-pressure ice is lower than ${\sim} 10^{15}$ Pa s \citep{rappaport2008can}. The viscosity of high-pressure ice is typically comparable to or greater than this value when the temperature of the ice is ${\sim}10 \%$ lower than the melting temperature \citep{Durha-etal:1998}, a reasonable assumption at the ocean floor. 

Since resonantly excited internal gravity waves can produce the required dynamic gravity, the main challenge for resonant stratification is how to establish a long-lived tidal resonance in an icy satellite. The possibility of resonant tidal excitation is suggested by the tendency of Titan's ocean to freeze due to secular cooling, which imposes a continuous evolution on the frequency of ocean waves by a change in the stratification profile of the ocean. At some point in Titan's freezing history, the normal frequency of ocean waves crosses the orbital frequency. After first being encountered, a tidal resonance can be sustained over long geological timescales provided that a stable fixed point is reached between orbital and interior evolution, analogously to ideas previously applied to tidal resonant locking in Jupiter and Saturn \citep{fuller2016resonance,luan2018cassini,lainey2020resonance,idini2022gravitational}. The onset of resonant ocean waves enhances Titan's tidal heating in the ocean \citep{tyler2011tidal,tyler2014comparative,rovira2019tidally}, with the additional heating slowing or halting the freezing of the ocean. 

\section{Basic Equations\label{sec:equations}}

The tidal Love number $k_{\ell m}$ represents a normalization of the gravitational tidal response by the gravitational excitation,
\begin{equation}
    k_{\ell m}= \left(\frac{\phi'_{\ell m}}{\phi^e_{\ell m}}\right)_{r=R} \textnormal{,}
    \label{eq:love_k}
\end{equation}
where the eccentric gravitational excitation $\phi^e_{\ell m}$ is derived in Appendix \ref{app:forcing} and the tidal response $\phi'_{\ell m}$ due to eccentricity tides is calculated numerically and analytically in this section in the context of various models. \added{We concentrate in the $\ell=m=2$ Love number $k_2$, which matches the spherical harmonic of the Cassini observation. We do not discuss obliquity tides in this manuscript because they excite $m=1$ spherical harmonics that do not contribute to the Cassini $k_2$ observation.}

\subsection{Equilibrium tide}
\added{Here we reproduce previously known results of the tidal response of icy satellites with oceans using simple principles. Our objective is to illuminate the effects contributing to the amplitude of the equilibrium tide and to produce an estimate for Titan within a few percent of accuracy. We start by deriving the hydrostatic $k_2$ in a homogeneous fluid body. This derivation shows how all the tidal gravity in $k_2$ emerges from a thin layer of displaced fluid near the surface. However, Titan strongly departs from this estimate because its density profile is not homogeneous and its tidal response is not fully hydrostatic. Our second derivation introduces the effects of a nonhomogeneous density profile and an inner core that responds rigidly to diurnal tides (i.e., tidal frequency $\omega=\omega_s$) rather than hydrostatically. Finally, our last derivation shows how an elastic ice shell covering the ocean reduces the tidal response and provides a rough estimate that agrees with more sophisticated modeling. After considering all these effects acting jointly, the result is what we consider the diurnal equilibrium tide of Titan $k_2\approx0.42$; a tide restored purely by static forces where inertial forces are neglected (i.e., no dynamics).}

\subsubsection{Hydrostatic tides in a homogeneous fluid body}

We first revisit the classical problem of tides in a homogeneous incompressible body that satisfies hydrostatic equilibrium using basic principles. The linearized equation for the conservation of momentum is
\begin{equation}
    \frac{p'}{\rho} = \tilde{\phi}'
    \label{eq:hydrostatic}
    \textnormal{,}
\end{equation}
The tidal response of the icy satellite produces adiabatic perturbations represented in the gravitational potential $\phi'$ and pressure $p'$. In this expression, the potential of the gravitational pull $\phi^e$ and the tidal gravitational potential $\phi'$ are combined into $\tilde{\phi}'=  \phi^e+\phi'$ for analytical simplicity.  

In an incompressible body, the density $\rho$ remains constant regardless of forcing and the only change in the gravity field comes about from the radial displacement $\xi_r$ of the surface. This diplacement is typically small compared to the radius $R$. Thus, we can calculate the gravity field from the integration over the volume of a thin shell of fluid with thickness $\xi_r$,
\begin{equation}
    \phi'_{\ell m} =\frac{4\pi \mathcal{G}\rho}{(2\ell +1)}\frac{R^{\ell+2}}{r^{\ell+1}}\xi_{r\ell m} = \frac{3}{(2\ell +1)}\left(\frac{R}{r}\right)^{\ell+1}g\xi_{r\ell m}\textnormal{,}
    \label{eq:tidal_surf}
\end{equation}
where $g$ is the surface gravity acceleration calculated in hydrostatic equilibrium $g=4\pi\mathcal{G}\rho R/3$, and $\mathcal{G}$ the gravitational constant. 

The boundary condition on the surface indicates that the fluid is free from external pressure (i.e., $\delta p =0$). This statement translates into turning the tidal displacement into the sole source for the pressure perturbation, formally expressed as
\begin{equation}
    \frac{p'}{\rho} = -\frac{\xi}{\rho} \frac{\partial p}{\partial r} = g\xi_r \textnormal{.}
\end{equation}

Next, we can go back to equation~(\ref{eq:hydrostatic}) and obtain an expression for the tidal forcing,
\begin{equation}
    \phi^e_{\ell m} = \frac{p'_{\ell m}}{\rho} - \phi'_{\ell m} =\left(1 - \frac{3}{2\ell+1}\left(\frac{R}{r}\right)^{\ell+1}\right) g\xi_{r\ell m}\textnormal{.}
\end{equation}

We evaluate the tidal response and forcing at the surface ($r=R$) to obtain the tidal Love number from the ratio between them,
\begin{equation}
    k_{\ell m} = \frac{3}{2(\ell-1)}
    \textnormal{,}
\end{equation}
which agrees with the classical result $k_2=1.5$ of the fluid Love number of a uniform density body in hydrostatic equilibrium \citep[e.g.]{munk1977rotation}.

\subsubsection{A global ocean with a rigid ocean floor \label{sec:k2core}}

We now consider a body with a rigid core of density $\rho_c$ and radius $R_c$ overlaid by an ocean of density $\rho$ by introducing vertical differentiation in the density profile. The main change compared to the homogeneous case is a change in the expression for the surface gravity acceleration $g=4\pi\mathcal{G}\bar{\rho} R/3$, where $\bar{\rho}$ is the body's mean density. The tidal gravity field still emerges solely from the radial displacement $\xi_{r\ell m}$ of a thin region near the surface, 
\begin{equation}
    \phi'_{\ell m} = \frac{3}{(2\ell +1)}\left(\frac{\rho}{\bar{\rho}}\right)\left(\frac{R}{r}\right)^{\ell+1}g\xi_{r\ell m}\textnormal{.}
\end{equation}
Following the same steps as before, the resulting tidal Love number is independent of core properties,
\begin{equation}
    k_{\ell m} = 
   \frac{3}{\left(2\ell +1\right)\bar{\rho}/\rho-3}\textnormal{.}
   \label{eq:k2}
\end{equation}
This equation yields $k_2=0.468$ when using Titan's mean density $\bar{\rho}\approx 1.88$ g~cm$^{-3}$ and $\rho=1$~g~cm$^{-3}$. Equation~(\ref{eq:k2}) reproduces numerical results (within 5\%) obtained previously for Europa when the icy shell is ignored \citep{moore2000tidal}, namely $k_2\approx0.249$ when using $\bar{\rho}\approx 3.01$ g~cm$^{-3}$. Analogous versions of Equation (\ref{eq:k2}) can be found in the literature when following an alternative derivation \citep[e.g.]{dermott1979shapes,murray1999solar,beuthe2015tides}.

\subsubsection{Ice shell thickness \label{sec:shell}}

An elastic ice shell constrains the radial displacement of the ocean that produces the tidal response registered in $k_2$. The weight of the ocean trying to reach an equipotential surface is balanced by the resistance of the ice shell to deformation \citep[e.g.]{Kamat-etal:2015}. Here we provide a simple argument to evaluate the role that the ice shell thickness plays in reducing the amplitude of the tidal response. 

Consider a global ocean of density $\rho$ that is trying to reach the equipotential surface defined by the radial displacement $\xi_{req}$. The pressure that the weight of the ocean exerts on the base of the ice shell can be expressed by 
\begin{equation}
p\sim \rho g (\xi_{req}-\xi_r) \textnormal{,}
\end{equation}
where $g$ is the gravity acceleration and $\xi_r$ is the radial tidal displacement. The shear stress $\tau$ on the ice shell can be expressed as
\begin{equation}
    \tau \sim \frac{\mu \xi_r}{R}\textnormal{,}
\end{equation}
where $\mu$ is the shear modulus of the ice shell. 

Next, we consider the equilibrium of forces over a meridional section dividing the satellite into two hemispheres. The ocean pressure integrated over the projected area of the ice shell base must be balanced by the shear stress integrated over the section of the ice shell, namely
\begin{equation}
    \pi R^2\rho g (\xi_{req}-\xi_r) \sim 2\pi d \mu \xi_r \textnormal{,}
\end{equation}
where $d$ is the ice shell thickness and $R$ is the icy satellite radius. We can rearrange the previous expression to discover the fractional change in $k_2$ due to a rigid ice shell,
\begin{equation}
    \Delta k_2 \sim\frac{\xi_r}{\xi_{req}} -1 \sim -\frac{2d\mu}{R^2\rho g}\textnormal{,}
    \label{eq:dk2}
\end{equation}
which represents a competition between elastic and gravitational energy (see also  equation~(11) in \cite{GoldrMitch:2010}). \added{An alternative derivation of equation (\ref{eq:dk2}), including all the correct numerical factors, can be found in \cite{beuthe2015tides}.} We may use Titan's radius $R\approx 2575$ km, surface gravity acceleration $g \approx 135$ cm s$^{-2}$, ocean density $\rho\approx 1$ g cm$^{-3}$, and shear modulus of ice I and methane clathrates $\mu \sim 4$ GPa, and obtain
\begin{equation}
    \Delta k_2 \sim -10 \textnormal{\%}\left(\frac{d}{100 \textnormal{ km}}\right)\textnormal{.}
\end{equation}
An ice shell thickness $d\sim100$ km balances radiogenic heating with heat conduction (Appendix~\ref{app:energy}) and produces $k_2\approx0.42$ (the hydrostatic response without an ice shell is $k_2=0.468$). We use this $k_2$ value as the reference hydrostatic tidal response when computing the fractional change $\Delta k_2$ due to various effects, which is compatible with previous estimates for Titan models without salinity \citep{rappaport2008can}. The result is valid for a shell made of ice or methane clathrates given that the elastic modulus is similar in both cases. 

\added{Instead of being purely elastic, the ice shell covering the ocean can be viscous near the melting temperature. Viscous ice can in principle flow at certain timescale and reduce the resistance that the ice shell imposes on the tidally excited ocean motion. In practice, this effect is negligible at the tidal timescale ${\sim} 16$ days unless large portions of the $d\sim 100$ km ice shell thickness is in convection and thus have relatively low viscosity (${\sim}10^{14}$ Pa s). The compensated long-wavelength topography of Titan suggests that the ice shell is unlikely to be convecting \citep{nimmo2010shell}.}

\subsection{Dynamical tides\label{sec:dyn_tides}}

We now solve the problem of a rotating ocean world with a stratified ocean subjected to the full action of the Coriolis effect. \added{The equation of conservation of momentum and continuity are}
\begin{equation}
    \partial_t \bm{v} + 2{\bf\Omega} \times \bm{v} = \nabla \tilde{\phi}-\frac{ \nabla p}{\rho} - \gamma \bm{v}
     \textnormal{,}
     \label{eq:rawmomentum}
\end{equation}
\begin{equation}
     \nabla\cdot {\bm v} = 0\textnormal{,}
     \label{eq1:continuity}
\end{equation}
 where $\bm v$ is the \added{3D incompressible} tidal flow, $\bm \Omega$ is the icy satellite spin vector, $\rho$ is the ocean density, $p$ is pressure, and $\gamma$ is the linear damping coefficient that represents dissipation in the ocean. In this paper, we consider at all times that Titan is in synchronous rotation (i.e. $\Omega=\omega_s$).

\added{We derive the linearized form of the conservation of momentum equation by introducing linear perturbations of the form
\begin{eqnarray}
    \tilde{\phi} \approx \phi_0 + \tilde{\phi}'\textnormal{,}\nonumber\\
    p \approx p_0 + p'\nonumber\\
    \rho \approx \rho_0 + \rho'\textnormal{,}
\end{eqnarray}
where quantities with the zero subindex indicate the background state and primed quantities indicate perturbations due to tides. In the following, we drop the zero subindex for convenience, thus all nonprimed quantities indicate the background state.} The resulting linearized equation of conservation of momentum is
\begin{equation}
     \partial_t \bm{v} + 2{\bf\Omega} \times \bm{v} = \nabla \tilde{\phi}'-\frac{ \nabla p'}{\rho} + \frac{\rho'}{\rho^2}\nabla p - \gamma \bm{v}
     \textnormal{.}
     \label{eq:momentum_gmode}
\end{equation}

The timescale of tidal motion of ${\sim}15$~days is orders of magnitude shorter than the timescale required to transport heat by diffusion or convection. Tides are consequently adiabatic and the associated Lagrangian tidal perturbations in density and pressure satisfy
\begin{equation}
    \frac{\delta p}{p} = \Gamma \frac{\delta\rho}{\rho}\textnormal{,}
\end{equation}
where $\Gamma$ is the adiabatic index. We can relate the Lagrangian perturbations to the Eulerian perturbations in equation~(\ref{eq:momentum_gmode}) using the definitions
\begin{equation}
    \delta p = p' + {\bm \xi}\cdot \nabla p\textnormal{,}
\end{equation}
\begin{equation}
    \delta \rho = \rho' + {\bm \xi}\cdot \nabla \rho \textnormal{,}
    \label{eq:Lrho}
\end{equation}
where ${\bm \xi}$ is the tidal displacement. Moving forward, we put together the latter equations to arrive to
\begin{equation}
     \frac{\rho'}{\rho} = \frac{p'}{\Gamma p}  + \bm{\xi} \cdot \frac{ \bm{N}^2}{g}\textnormal{,}
     \label{eq:Ngamma}
\end{equation}
where $N^2$ is the Brunt-Vaisala (BV) frequency defined in general by
\begin{equation}
    \bm{N}^2 = g\left( \frac{\nabla p}{\Gamma p} - \frac{\nabla\rho}{\rho}\right)\textnormal{.}
\end{equation}
The BV frequency represents the salinity stratification of the ocean. Here the only relevant component in $N^2$ is the radial direction, given that we consider the background state to be spherically symmetric. When $\hat{\bf r}N^2>0$, the ocean is vertically stratified. A stratified ocean parcel will develop static stability once displaced out of its equilibrium position. 

According to equation~(\ref{eq:Ngamma}), an Eulerian change in ocean density may emerge from either the adiabatic response of the ocean fluid (first term in the right-hand side) or the buoyancy of the stratified ocean parcel (second term in the right-hand side). In the incompressible approximation used here, Eulerian perturbations to density uniquely emerge from the buoyancy of the stratified ocean parcel. This results from the adiabatic index $\Gamma \equiv (\partial\log p/\partial\log\rho)_S$ tending to infinity: the expected changes in pressure keep density unperturbed. As a consequence, the BV frequency reduces to
\begin{equation}
    \bm{N}^2 = -g\frac{\partial_r\rho}{\rho} \hat{\bf r}
    \textnormal{,}
    \label{eq:BV2}
\end{equation}
where the radial variation in density comes from the addition of a small amount of salts organized in a vertical salinity gradient. As we can see from equation~(\ref{eq:BV2}), stratification permits local density perturbations in the ocean regardless of its incompressibility. 

Equation~(\ref{eq:BV2}) shows a direct relationship between stratification in $N^2$ and the salinity gradient in $\partial_r\rho$. When we consider constant stratification throughout the ocean, we can write $\partial_r\rho=\Delta\rho/H$, where $\Delta\rho$ is the change in density between top and bottom of the ocean. Our models have constant density, thus the $\Delta\rho$ represents a virtual change in density due to the addition of salts. The concentration of added salts $S$ is described in g of salts per kg of water (g/kg). 

Next, we rewrite the linearized momentum and continuity equations as
\begin{equation}
     \omega \tilde{\omega} {\bm \xi} + i2\omega{\bf\Omega} \times {\bm \xi} - \left(\bm{N}^2\cdot {\bm \xi}\right) \hat{\bf r} =  -\nabla\psi\textnormal{,}
     \label{eq:momentum_gmode2}
\end{equation}
\begin{equation}
   \psi = \tilde{\phi}'- \frac{ p'}{\rho} \textnormal{,}
\end{equation}
\begin{equation}
     \nabla\cdot {\bm \xi} = 0\textnormal{,}
     \label{eq1:continuity}
\end{equation}
where $\tilde{\omega} = \omega + i\gamma$ is a complex frequency that accounts for tidal dissipation, $\omega=\omega_s$ is the tidal frequency of eccentricity tides, and the tidal displacement $\bm{\xi}$ is periodic with a time dependency $\propto e^{-i\omega t}$. \added{We have used $\bm{v}=\partial_t {\bm \xi}= -i\omega{\bm\xi}$}. This set of equations is traditionally known as the Boussinesq approximation. 

A typical method of solution of equation~(\ref{eq:momentum_gmode2}) involves applying the curl to reach the vorticity equation \citep{rieutord1987linear}
\begin{equation}
 \nabla\times\left(2\bm{\Omega}\times\bm{\xi}\right)  +i\tilde{\omega}\nabla\times\bm{\xi} + \frac{1}{i\omega}\nabla\times\left(\left(\bm{N}^2\cdot {\bm \xi}\right) \hat{\bf r}\right) = 0\textnormal{.}
 \label{eq:vort}
\end{equation}
A rigid core implies that waves produce no radial displacement near the bottom of the ocean. We set the rigid boundary condition at the ocean bottom to
\begin{equation}
    \bm{\xi} \cdot \hat{\bf r} = 0
    \textnormal{.}
\end{equation}
The free boundary at the ocean top prescribes a zero Lagrangian perturbation of pressure (e.g., \cite{goodman2009dynamical}), 
\begin{equation}
    \delta p = p' + \bm{\xi}\cdot\nabla p= 0 
    \textnormal{,}
\end{equation}
which leads to
\begin{equation}
    \left( g - \frac{4\pi\mathcal{G}\rho R}{2\ell+1}\right)\bm{\xi} \cdot \hat{\bf r} + \psi = \phi^e_{\ell m}\textnormal{.}
\end{equation}

The tidal displacement field $\bm{\xi}$ is the only quantity to be determined, which is forced by the ocean top boundary condition on the radial component of the displacement. 
Previous work \citep[e.g.]{rovira2019tidally} typically sets $\bm{\xi} \cdot \hat{\bf r}$ at the surface to be equal to the equilibrium tide \added{radial} displacement (i.e., no-slip boundary conditions) and/or neglects the self gravitation of the tide (e.g., Cowling approximation). Given that we are interested in precise tidal gravity estimates, we have retained the self gravitation of the equilibrium tide and allowed the ocean top to displace dynamically beyond the equilibrium tide, \added{relaxing the no-slip assumption}. No-slip boundary conditions explicitly set the dynamical part of $\bm{\xi} \cdot \hat{\bf r}$ to zero at the surface, removing  by definition the dynamical enhancement of $k_2$ that we calculate here.

The resulting set of equations is an infinite system of equations coupled in degree by the Coriolis effect \citep[e.g.]{rieutord1987linear,rieutord1997inertial,lockitch1999r,rieutord2001inertial,ogilvielin2004tidal,ogilvie2009tidal,rovira2019tidally,idini2021dynamical}. We solve this system by the traditional method of projection onto vectorial spherical harmonics (Appendix~\ref{app:proj}), followed by a pseudo-spectral discretization on the radial functions based on the analytically-tractable Chebyshev polynomials and Gauss–Lobatto collocation points \citep[e.g.]{boyd2001chebyshev}. We truncate the infinite set of equations at degree $\ell_{max}=100$ and use $N_{max}=100$ Chebyshev polynomials to represent each radial function at each degree $\ell\leq\ell_{max}$. The Love number can then be obtained from the tidal displacement evaluated at the surface (equation~(\ref{eq:tidal_surf})), which is the only source of tidal gravity in a homogeneous ocean.


\subsection{Tidal heating rate in the ocean}

The heating rate in the ocean can be fully determined by volume integration \added{of the work done by the dissipative force in the equation of motion (equation (\ref{eq:rawmomentum})). This work per unit volume follows}
\begin{equation}
    dw = -\rho\gamma \bm{v}^*\cdot d\bm{l}\textnormal{,}
\end{equation}
\added{where $d\bm{l}$ is an infinitesimal line element along a fluid parcel motion. We can use $d\bm{l}=\bm{v}dt$ and integrate $\dot{w}$ across the ocean volume to obtain the time averaged ocean heating rate \citep[e.g.]{chen2014tidal,rovira2023thin}}
\begin{equation}
    \dot{E}= \rho\gamma \omega^2 \int_V   \bm{\xi}^*\cdot\bm{\xi}dV\textnormal{.}
    \label{eq:heating}
\end{equation}
The linear damping coefficient $\gamma$ is unconstrained in icy worlds with global oceans. The range of possible estimates spans from $\gamma\sim10^{-11}$ s$^{-1}$ on Enceladus to the better constrained $\gamma\sim10^{-5}$ s$^{-1}$ on Earth \citep{matsuyama2018ocean}. The projection of equation (\ref{eq:heating}) onto vectorial spherical harmonics is shown in Appendix~\ref{app:proj}.

\section{Numerical results}

\subsection{The predicted tidal response $k_2$ as a function of ocean structure}

We use perturbation theory and numerical methods to calculate the dynamic gravity produced by resonant stratification in a stratified and rotating ocean (Section~\ref{sec:equations}). We concentrate on fractional changes $\Delta k_2$ to the non-resonant hydrostatic $k_2\approx0.42$ obtained in a pure water ocean with a $d\sim100$ km ice shell resting on top. Our method of solution considers the Coriolis force in full and avoids the thin shell approximation typically used in tidal studies of icy satellites \citep{beuthe2016crustal,matsuyama2018ocean,rovira2023thin}. The additional effort of relaxing the thin-shell approximation allows us to study the dynamic gravity of internal gravity waves that result from the mixing of rotational and stratification effects. We simplify the structure of stratification by assuming a constant $N^2$ throughout the ocean (equation (\ref{eq:BV2})), which translates into a linear increase in salt concentration with depth starting from zero at the ocean surface. More complicated salinity distributions are in principle possible and lead to additional uncertainty in the inferred ocean salinity structure.

\begin{figure}[ht!]
    \centering
    \plotone{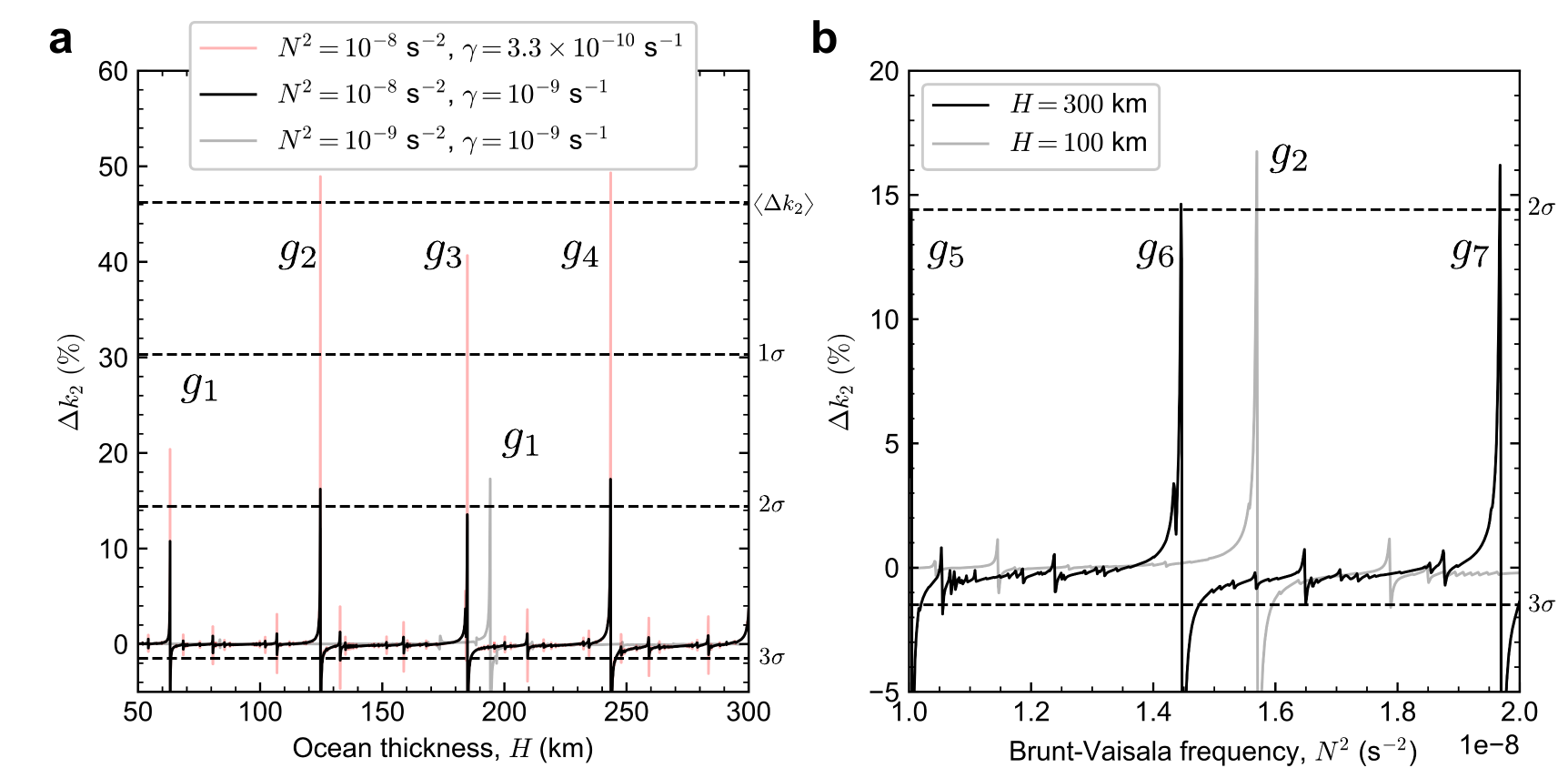}
    \caption{Titan's $k_2$ enhancement (fractional correction $\Delta k_2$) from dynamic gravity as a function of (a) ocean thickness $H$ and (b) the Brunt-Vaisala frequency $N^2$. Peaks represent resonances with ocean normal modes. The ocean top salinity is zero and increases linearly with depth. The ocean bottom salinity ranges $0.8-1.5$ g/kg ($H=100$ km) and $2.3-4.4$ g/kg ($H=300$ km) for the range of $N^2$ shown in (b). Frictional dissipation in (b) is $\gamma=10^{-9}~\rm s^{-1}$.  The reference hydrostatic Love number is $k_2=0.42$, a pure water ocean with an elastic ice shell with thickness $d\sim100$ km. The tidal frequency is equal to the rotational frequency and the orbital frequency, $\omega=\Omega=\omega_s=2\pi/T_{orb}$, where $T_{orb}=15.945$ days is Titan's orbital period. \label{fig:k2_resonant}}
\end{figure}

Our models show that resonant stratification can explain the $k_2$ enhancement observed by Cassini. We observe an enhancement $\Delta k_2$ beyond $+15\%$ when overtones composed of internal gravity waves are resonantly excited by eccentricity tides (Fig.~\ref{fig:k2_resonant}). This brings $k_2$ from $3\sigma$ to $2\sigma$ away from the mean value of the observation (Fig.~\ref{fig:k2_resonant}). In this case, we have used the conservative linear damping $\gamma=10^{-9}$ s$^{-1}$, but our models predict a resonant $\Delta k_2$ beyond $+45\%$ when using a still realistic $\gamma=3.3\times10^{-10}$ s$^{-1}$ (Fig.~\ref{fig:k2_resonant}), an enhancement that puts $k_2$ at the mean value of the observation at the saturation point of the resonance. Resonances occur at various $H$ and $N^2$, preventing us from identifying a unique ocean thickness and stratification profile based solely on $k_2$. In resonant stratification, the salt concentration near the ocean floor can be as low as $<5$ g/kg in the simplified model we use here (Fig.~\ref{fig:k2_resonant}; equation (\ref{eq:BV2})). The mean salinity can then be less than that for the oceans of Earth or Enceladus \citep{postberg2009sodium}, while still producing a resonant response.

\begin{figure}[ht!]
    \centering
    \plotone{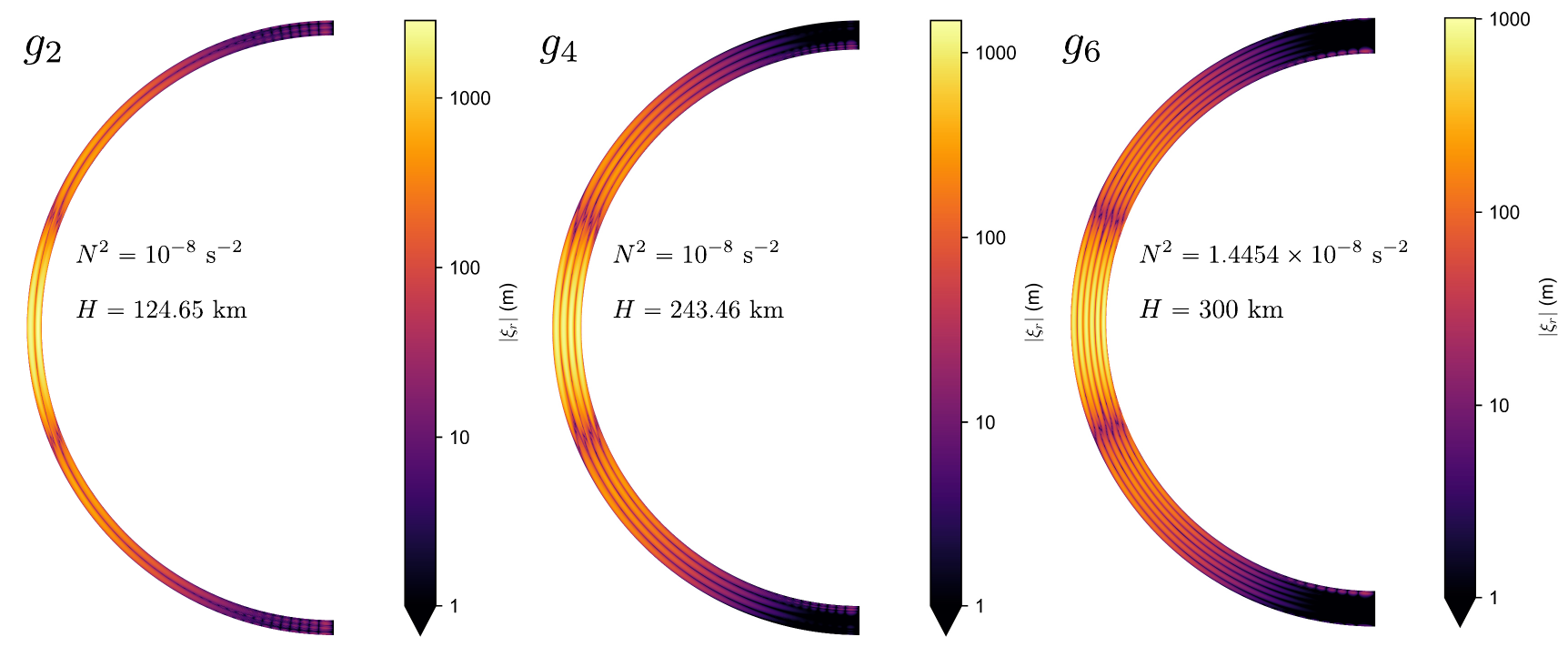}
    \caption{Meridional cross section of the radial displacement on Titan's ocean due to tides for selected internal gravity wave normal modes of increasing radial order shown in Fig.~\ref{fig:k2_resonant}.  Frictional dissipation is $\gamma=10^{-9}~\rm s^{-1}$.  \label{fig:radial}}
\end{figure}

\begin{figure}[ht!]
    \centering
    \plotone{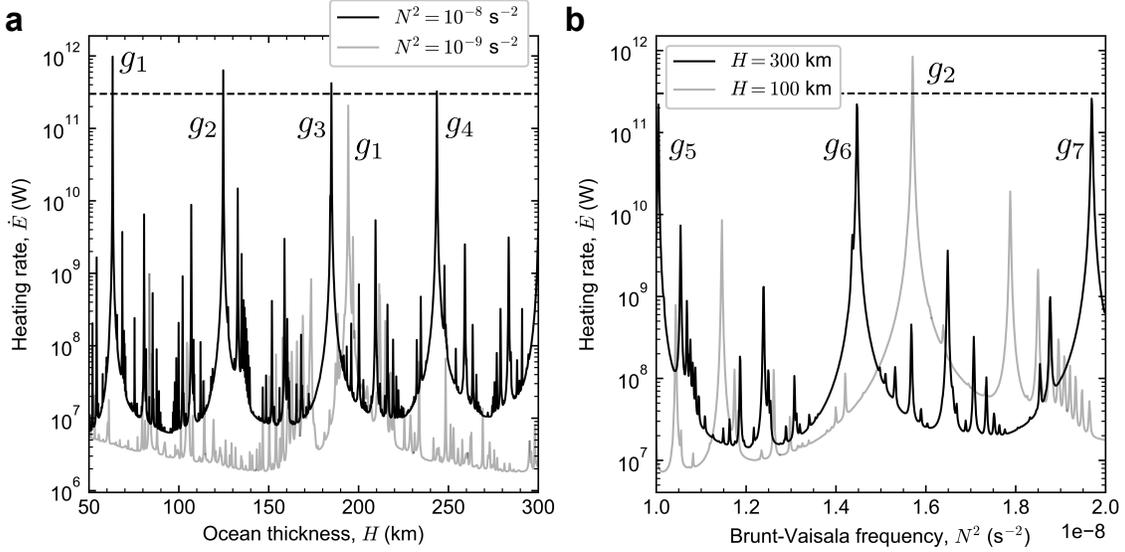}
    \caption{Heating rate produced by tidally excited internal gravity waves in the stratified ocean, as a function of (a) ocean thickness $H$ and (b) Brunt-Vaisala frequency $N^2$. The dashed line indicates Titan's radiogenic heating $\dot{E}_{int}\sim3\times10^{11}$ W (see Apprendix~\ref{app:energy}). The frictional damping is $\gamma=10^{-9}$ s$^{-1}$. Overtones of g-modes are labeled with a subindex representing the mode radial order $n$. The smaller peaks without a label represent dissipation from resonant inertial wave modes/attractors not discussed here (see \cite{rovira2019tidally}, e.g.). \label{fig:k2}}
\end{figure}

As a general rule, internal gravity waves become resonant when the nodes in the radial displacement of the ocean wave perfectly fit the thickness of the stratified ocean cavity (Fig.~\ref{fig:radial}). We can achieve this resonant fit by adjusting $N^2$ and changing the radial wavelength of internal gravity waves, or by adjusting the thickness of the stratified ocean $H$. The number of radial nodes is directly proportional to overtone order and inversely proportional to mode frequency, with lower frequency internal gravity waves having more radial nodes \citep{unno1979nonradial}. Increasing the strength of stratification $N^2$ shifts the spectrum of g-modes toward high frequency and allows higher order gravity modes to become resonant with the fixed orbital frequency (Fig.~\ref{fig:k2}). 

\subsection{Surface radial displacement and interior nonlinear wave breaking}

The dynamical enhancement $\Delta k_2$ emerges from a tens-of-meter enhancement on the radial displacement of the ocean surface. \added{In our models, this fractional enhancement to ocean surface radial displacement $\Delta h_2$ equals $\Delta k_2$ for any given combination of ocean parameters. We derive this result from combining equations (\ref{eq:love_k}), (\ref{eq:tidal_surf}), and the definition of the tidal Love number $h_{\ell m}$,} 
\begin{equation}
    h_{\ell m}= g\left(\frac{\xi_{r\ell m}}{\phi^e_{\ell m}}\right)_{r=R} \textnormal{,}
    \label{eq:love_h}
\end{equation}
resulting in the relationship
\begin{equation}
    h_{\ell m} = \left(\frac{2\ell+1}{3}\right)\left(\frac{\bar{\rho}}{\rho}\right)k_{\ell m}\textnormal{.}
    \label{eq:love_kvsh}
\end{equation}
\added{A direct inspection of equation (\ref{eq:love_kvsh}) leads to $\Delta h_2=\Delta k_2$ after substitution of the $h_{\ell m}$ and $k_{\ell m}$ that include a base value and a fractional correction. The Love number $h_{\ell m}$ describes the radial displacement of the ocean surface as a function of the equilibrium tide. A $h_{\ell m}=1$ implies that the surface follows the shape of the equilibrium tide when the self gravity of the tide is ignored. In a homogeneous fluid body, the average density is  $\bar{\rho}=\rho$ and equation (\ref{eq:love_kvsh}) recovers the classical result $h_2 = 5/2$. In the case of simple Europa interior models, equation (\ref{eq:love_kvsh}) reproduces previous numerical results for $h_2$ when the core/mantle are approximately rigid \citep{moore2000tidal}. In general, equation (\ref{eq:love_kvsh}) is valid for simple models where the tidal gravity originates entirely from the radial displacement of the surface, as is approximately the case for icy satellites with global oceans overlying roughly rigid rocky interiors.}

\added{When we consider that Titan's equilibrium tide produces a surface displacement $|\xi_r|\approx 26$ m, the dynamical enhancement from resonant stratification shown in Fig.~\ref{fig:radial} leads to tides with surface displacement below $|\xi_r|\lessapprox 40$ m in the most dramatic case. We can calculate Titan's equilibrium tide Love number $h_{\ell m}$ using equations (\ref{eq:k2}) and (\ref{eq:love_h}),
\begin{equation}
    h_{\ell m} = \frac{(2\ell +1)\bar{\rho}/\rho}{(2\ell+1)\bar{\rho}/\rho -3}\textnormal{.}
\end{equation}
Titan responds to Saturn's gravity with an equilibrium tide $h_2\approx1.47$, which is reduced by $-10\%$ when a $d\sim100$ km ice shell is included (Section~\ref{sec:shell}). We can obtain the radial displacement of Titan's equilibrium tide from equations (\ref{eq:love_h}), (\ref{eq:U}), and (\ref{eq:forcing})},
\begin{equation}
    |\xi_r| \approx \frac{7eh_2U_{22}}{g} = 7eh_2\left(\frac{3\pi}{10}\right)^{1/2} \left(\frac{MR^4}{m_sa^3}\right)\approx26\textnormal{ m,}
\end{equation}
where $M$ is Saturn's mass and $m_s$ is Titan's mass.

Below the ocean surface, resonant internal gravity waves produce negligible gravity but attain a radial displacement $|\xi_r|\sim 1$~km in the case of $\gamma=10^{-9}$ s$^{-1}$ (Fig.~\ref{fig:radial}). A lower dissipation $\gamma$ produces even larger resonant amplitudes of tidal motion interior to the ocean (Fig.~\ref{fig:g6}), with $\gamma=10^{-10}$~s$^{-1}$ reaching $|\xi_r|\sim 10$~km. This large radial displacement is accompanied by a horizontal displacement that is typically $\xi_\perp/\xi_r\sim n R/H \sim 50$ (equation~(\ref{eq:cont})) and allows the flow to preserve continuity in an incompressible fluid. Despite the large tidal displacement, our models of resonant tidal excitation remain far from experiencing nonlinear wave breaking when $\gamma\gtrsim 10^{-10}$~s$^{-1}$, as stipulated in $|\xi k|\lesssim 1$, the typical criterion to avoid nonlinear wave breaking, where $k$ is the wavenumber. For the radial displacement, this criterion translates to $|\xi_r|\lesssim H/n$. In the case of the horizontal displacement, the criterion stipulates $\xi_\perp\lesssim \pi R/m$, where $m$ is the azimuthal order. All g-modes shown in Figs.~\ref{fig:k2_resonant} and \ref{fig:radial} avoid nonlinear wave breaking by at least one order of magnitude.

\subsection{Heating rate at saturation point}





Our models also show that resonant stratification produces enough heat to compete with the radiogenic heating generated by decaying isotopes inside solid Titan (Fig.~\ref{fig:k2}). This result is key to allow Titan to reach a stable fixed point that may prolong the crossing of a resonance as the ocean freezes. When in thermal steady state, the heat transported across the ice shell must balance all interior heat sources. If heat is transported by conduction, we can write $d \sim (3\times10^{13})/\dot{E}_{int}$ km (Appendix~\ref{app:energy}), where $\dot{E}_{int}$ is the interior heating rate in W and $d$ is the ice shell thickness. When the ocean freezes until balance with radiogenic heating, $\dot{E}_{int}\sim 3\times10^{11}$ W (Appendix~\ref{app:energy}) and the ice shell thickness grows to $d\sim100$ km. In this scenario, the ocean plays a negligible role in heating the interior. Resonant stratification changes this picture by introducing an additional heating source when the ocean is still rapidly freezing from secular cooling. In resonant stratification with $\gamma = 10^{-9}$~s$^{-1}$, for example, we get a total $\dot{E}_{int}\sim 6\times10^{11}$ W at saturation point (Fig.~\ref{fig:k2}) and the ice shell thickness becomes $d\sim50$ km while in steady state with internal heat from radiogenic heating and ocean tidal heating combined. 

The model above is a simple conductive model, but it provides a plausible argument in favor of catching resonant stratification via ocean freezing. Thermal steady state cannot be attained when the ice shell thickness is very thin near at the onset of ice shell formation ($d\lesssim 50$ km for $\gamma=10^{-9}$~s$^{-1}$); the resonance saturates at peak heating rate (Fig.~\ref{fig:k2}) before producing enough heat to balance conduction across the thin ice shell. Ice shell growth by secular cooling stops at $d\sim100$ km, hence resonant stratification via ocean freezing must catch a resonance before then. This last requirement is not significantly changed when a high abundance of ammonia in the ocean is considered. Ammonia can lower the freezing temperature of the ocean to the eutectic at $T\sim180$ K if present in concentration ${\sim}15$~wt.$\%$ \citep{lunine1987clathrate,grasset1996cooling}, reducing the $d$ required for thermal steady state in half independently of whether resonant stratification is in place or not (see equation (\ref{eq:pureh20})).

\begin{figure}[ht!]
    \centering
    \plotone{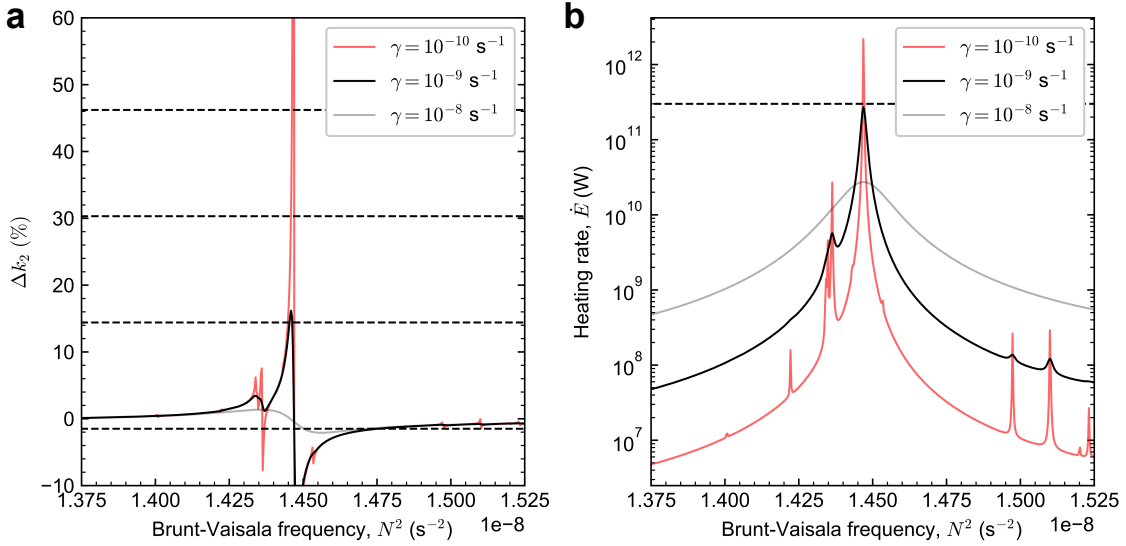}
    \caption{ Resonant dynamical gravity signal and heat production as a function of dissipation for g-mode $g_6$ ($\ell=m=2$ and $n=6$). Ocean thickness is $H=300$~km. The dashed lines are the same as in Fig.~\ref{fig:k2_resonant}, and Fig.~\ref{fig:k2}, respectively for (a) and (b).\label{fig:g6}}
\end{figure}

A lower $\gamma=10^{-10}$~s$^{-1}$ increases the heating rate at saturation point to an order of magnitude above Titan's radiogenic heating rate, and allows $k_2$ to grow higher  (Fig.~\ref{fig:g6}). A larger $\gamma$ leads to the damping of resonances, reducing the amplitude of the resonant $\Delta k_2$ and the resonant $\dot{E}$ (Fig.~\ref{fig:g6}). The hydrostatic flow is not much affected by $\gamma$, thus the heating rate is typically increased by a larger $\gamma$ when far from resonances. Our results are relevant in the range $\gamma=10^{-9}$~--~$10^{-10}$~s$^{-1}$; a $\gamma$ larger than this range damps $\Delta k_2$ below the signal registered by Cassini, whereas a $\gamma$ lower than the range results in nonlinear breaking of internal gravity waves, an effect not included in our models.

\section{Discussion}

\subsection{Approach to a long-lived resonance via ocean freezing \label{sec:freezing}}

\begin{figure}[ht!]
    \centering
    \plotone{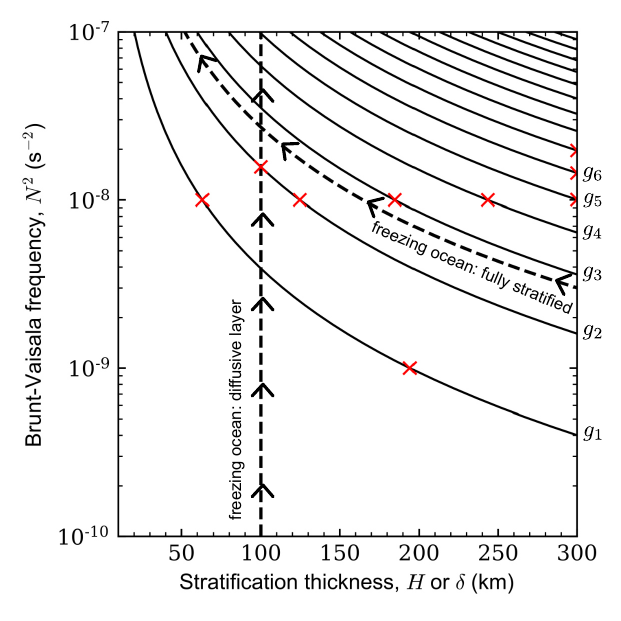}
    \caption{Model parameters that produce tidally excited resonances with internal gravity waves (solid lines; see equation (\ref{eq:gmodes_N})). The positive branch of the $\Delta k_2$ resonance extends toward the bottom-left of each solid line. The red crosses represent individual resonances identified in our numerical simulations and shown in Fig.~\ref{fig:radial}. The dashed lines represent simplified freezing trajectories for different models of ocean stratification. The fully stratified ocean model assumes $M_s \approx 8.3\times10^{21}$~g, which is equivalent to a $\delta\rho\sim0.001$~g~cm$^{-3}$ over an ocean thickness $H\sim200$~km. \label{fig:gmodes}}
\end{figure}

In addition to thermal steady state, ocean freezing requires an approach toward the resonance from the branch that produces a positive $\Delta k_2$ (Fig.~\ref{fig:k2_resonant}). Approaching the resonance from the negative $\Delta k_2$ branch would depart from the gravity enhancement required by Cassini. A simple freezing model of a fully stratified ocean fails to satisfy this criterion. In this freezing model, the ocean freezes leaving salts in the liquid phase and redistributing them into a linear salinity profile that conserves the total mass of salts $M_s$. Following equation~(\ref{eq:BV2}), the ocean stratification in this case follows
\begin{equation}
    N^2 \sim -\frac{gM_s}{2\pi\rho R^2H^2}\textnormal{.}
\end{equation}
Despite its attractive simplicity, a fully stratified ocean freezes following a trajectory that never converges toward a g-mode resonance, independently of the $M_s$ assumed (Fig.~\ref{fig:gmodes}). 

In an alternative freezing model, a diffusive layer of thickness $\delta$ is sandwiched between two layers of constant density, where the top layer has roughly no salinity and the bottom layer is high salinity. Internal gravity waves are excited in the stratified diffusive layer instead of the entire ocean. In this alternative case, the ocean freezes without changing the diffusive layer thickness $\delta$, but increasing the density contrast $\delta\rho$ and consequently increasing $N^2$ (equation~(\ref{eq:BV2})), according to
\begin{equation}
    N^2 \sim -\frac{gM_s}{4\pi\rho R^2h_b\delta}\textnormal{,}
    \label{eq:difflayer}
\end{equation}
where $h_b$ is the thickness of the bottom layer with high salinity. The freezing of an ocean with a diffusive layer (i.e., reducing $h_b$ in equation~(\ref{eq:difflayer}) at constant $\delta$) suggest the possibility of crossing the $\Delta k_2$ resonance from the positive branch (Fig.~\ref{fig:gmodes}). Further theoretical development is required to determine the effects of a diffusive layer in the $\Delta k_2$ results reported here.

A stable fixed point is then established once resonant stratification succeeds at reaching thermal steady state by halting ocean freezing with the additional heating rate provided by the resonance \citep{tyler2011tidal,tyler2014comparative}. The secular expansion of Titan's orbit pushes the orbital frequency away from resonance, reducing the heating rate and promoting further ice thickening until ocean waves tune again with the new orbital frequency. This can happen because ice shell thickening by secular cooling can keep up with the orbital evolution. For instance, the thermal adjustment timescale for a 100~km thick shell is $d^2/\pi^2 \kappa \approx30$~Myr; over this timescale the orbital frequency will have changed by 0.5\% (Section~\ref{sec:orbital}).  The alternative to a stable fixed point is that Titan has by chance encountered a resonance that will last until further orbital evolution pushes the orbital frequency out of resonance. 

\added{The two ocean freezing scenarios described above assume a radially isotropic distribution of salts that linearly increases with depth over certain radial distance, either the ocean thickness $H$ or the diffusive layer thickness $\delta$. More complicated distributions of salts are in principle possible. The distribution of salts can show lateral variations in the presence of nonuniform ice shell thickness due to alternating regions of melting and freezing below the ice shell \citep{ashkenazy2018dynamics,lobo2021pole,kang2022does,kang2023modulation}. Future investigations are required to better understand the impact of various distributions of salts in the dynamics of tidally excited internal gravity waves and the ocean freezing path that leads to a stable resonance.}

\subsection{Resonant stratification via orbital evolution \label{sec:orbital}}

\begin{figure}[ht!]
    \centering
    \plotone{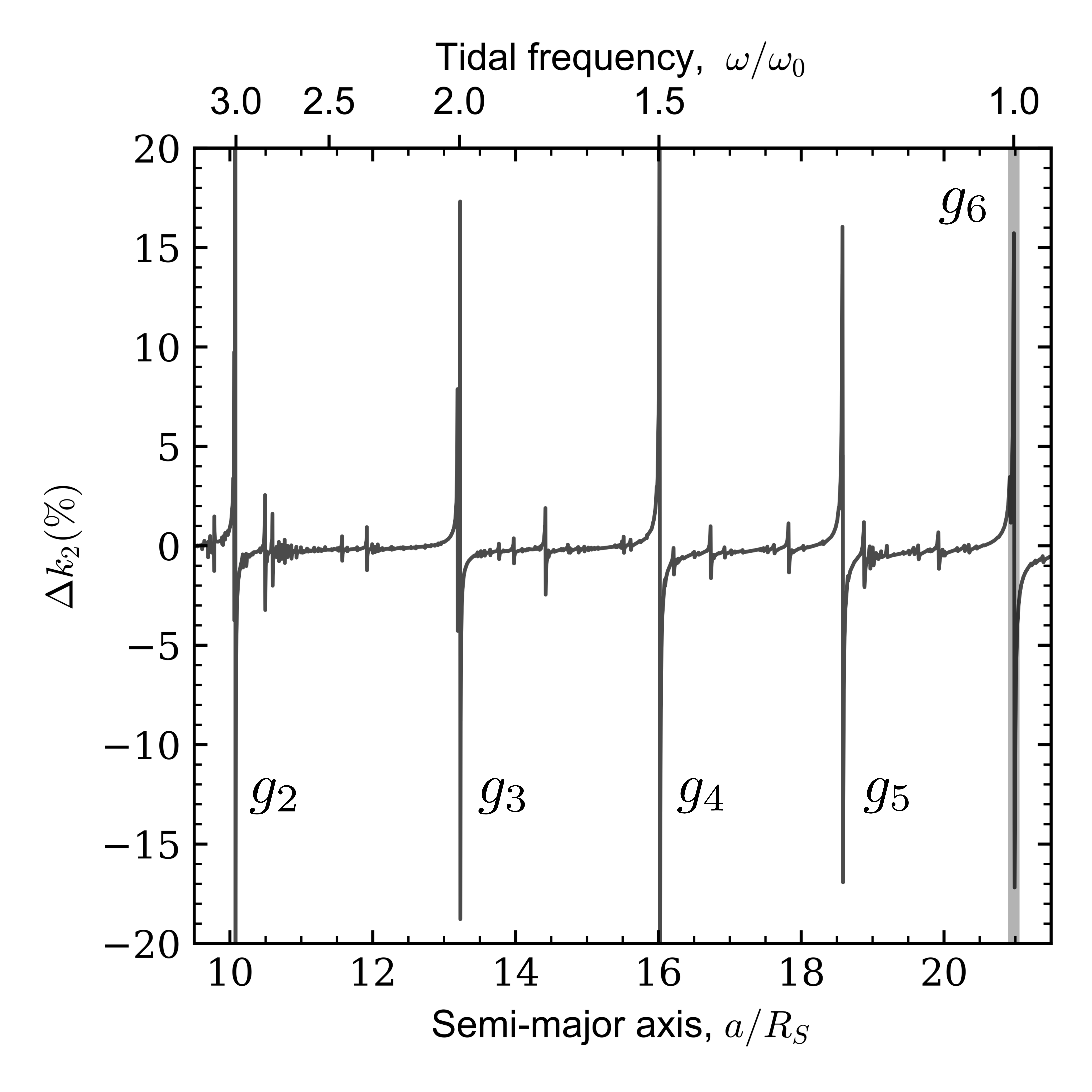}
    \caption{ Same as Fig.~\ref{fig:k2_resonant}, but as a function of orbital semi-major axis $a$. The shaded vertical line represents Titan's current semi-major axis $a=21 R_s$, where $R_s$ is Saturn's radius. The ocean is fixed to $H=300$~km and $N^2=1.4454\times10^{-8}$~s$^{-2}$ (see also Fig.~\ref{fig:radial}). The tidal frequency is equal to the orbital frequency and $\omega_0$ is Titan's current orbital frequency $\omega_s$. The eccentricity is fixed to Titan's current $e$ and we assume synchronous rotation at all times (i.e., $\Omega=\omega_s$). \label{fig:orbital}}
\end{figure}

In the absence of a stable fixed point, resonant stratification can still be established by orbital evolution. Titan orbits Saturn at a slower rate than Saturn's rotation, leading to outward migration from tidal torques that arise after tidal dissipation occurs inside Saturn \citep{lainey2020resonance}. This outward migration imposes a slow drift in the excitation frequency of eccentricity tides. At some point during orbital migration, the frequency of eccentricity tides can match the frequency of a g-mode overtone trapped in the stratified ocean, setting resonant stratification (Fig.~\ref{fig:orbital}). This mechanism operates without requiring any specific freezing history for Titan's ocean. The resulting resonance, however, is short lived. Continuous orbital migration will further break the resonance after pushing Titan through the resonance width $\delta a\sim 0.02 R_s$, where $R_s$ is Saturn's radius (Fig.~\ref{fig:orbital}) and we have assumed the reasonable $\gamma=10^{-9}$ s$^{-1}$ (i.e., the resonance width depends on dissipation, as shown in Fig.~\ref{fig:g6}). At Titan's current migration rate $\dot{a}/a\sim 10^{-10}$~yr$^{-1}$ \citep{lainey2020resonance}, the time it would take Titan to cover the $g_6$-mode resonance width would be $\delta a/\dot{a}\sim10$~Myr. 

The probability that Titan is simply passing over an ocean resonance is slim as a result of this fast orbital migration. The astrometric observation of Titan's migration \citep{lainey2020resonance} indicates that Titan's orbital period has increased ${\sim}3$ times by orbital expansion over the lifetime of the solar system $\lambda\sim4.5\times10^9$~yr, or
\begin{equation}
    \frac{\Delta P_s}{P_s} = \frac{3\Delta a}{2a} =\frac{3\lambda\dot{a}}{2a}\sim \frac{2}{3}\textnormal{,}
\end{equation}
where $P_s$ is Titan's current orbital period, and $\Delta P_s$ and $\Delta a$ represent orbital parameter changes on the timescale $\lambda$. The period spacing of g-modes (Appendix~\ref{app:modes}) allows us to estimate the number of g-modes that Titan crosses over the timescale $\lambda$,
\begin{equation}
    \#_g\sim 1+\frac{\Delta P_s}{\Delta P_g}\sim 1 + \frac{3\lambda\dot{a}P_s}{2a\Delta P_g}\textnormal{,}
\end{equation}
where $\Delta P_g$ is the g-mode spacing. Cassini registered Titan's enhanced $k_2$ at no particular time within Titan's interior evolution. The probability that Cassini observed a g-mode resonance motivated uniquely by orbital evolution (i.e., no stable fixed-point) is
\begin{equation}
    \mathcal{P}(\textnormal{g-mode}) \sim \#_g\left(\frac{\delta a}{\Delta a}\right) \sim \frac{\delta a}{a} \left(\frac{a}{\lambda\dot{a}} + \frac{3\pi}{\omega_s\Delta P_g} \right)\textnormal{.}
    \label{eq:prob}
\end{equation}
We obtain $\#_g\sim 5$ and $\mathcal{P}(\textnormal{g-mode}) \sim 1\%$ when using the reasonable ocean parameters in Fig.~\ref{fig:orbital}. Equation (\ref{eq:prob}) is only valid for $\#_g\gtrsim2$. The low $\mathcal{P}(\textnormal{g-mode})$ indicates that resonant stratification is unlikely to be a result of pure orbital evolution (i.e., no ocean freezing involved), yet not impossible. 

\subsection{A mildly salty ocean versus a heavy ocean}

\begin{figure}[ht!]
    \centering
    \plotone{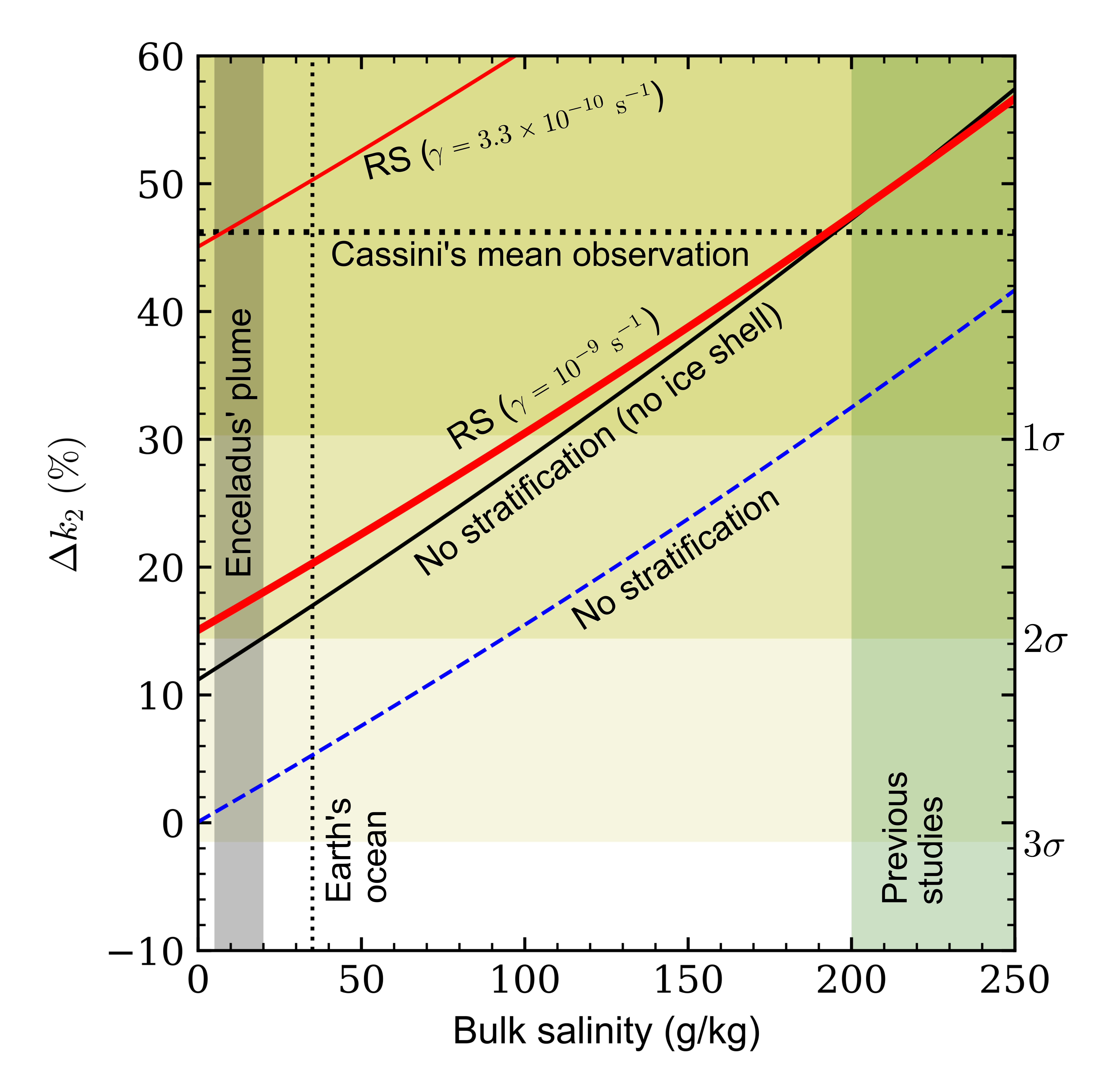}
    \caption{Titan's $k_2$ fractional enhancement as a function of ocean bulk salinity. For an ice shell thickness $d\sim100$ km and a given bulk salinity, resonant stratification (RS; solid red) can occur at various combinations of ocean thickness and B-.V. frequency, producing $\Delta k_2\approx +15\%$ ($\gamma=10^{-9}$~s$^{-1}$) and $\Delta k_2\approx +45\%$ ($\gamma=3.3\times10^{-10}$~s$^{-1}$) (see also Fig.~\ref{fig:radial}). The hydrostatic $k_2$ of a fully homogeneous ocean is shown in a dashed blue line (ice shell thickness $d\sim100$ km) and a solid black line (no ice shell). The background shade of yellow represents the fractional enhancement required to explain Cassini's observation at various confidence intervals \citep{durante2019titan}. The salinity in Enceladus' plume is from \cite{postberg2009sodium} and Titan previous studies are \cite{rappaport2008can,iess2012tides,mitri2014shape,durante2019titan}. \label{fig:dk2_mod}}
\end{figure}

When compared to previous studies, resonant stratification only requires a mild concentration of solute dissolved in the ocean to explain Titan's $k_2$ (Fig.~\ref{fig:dk2_mod}). With $\gamma=10^{-9}$~s$^{-1}$, resonant stratification yields a $k_2$ value 1$\sigma$ closer to the measured central value starting from the hydrostatic $k_2=0.42$. When $\gamma=3.3\times10^{-10}$~s$^{-1}$, the enhancement over the hydrostatic $k_2$ is 3$\sigma$, crossing the mean value of the Cassini $k_2$ observation at low salinity ($S<10$~g/kg). In the absence of resonant stratification, a heavy convective ocean requires a salt concentration that is on average $S\sim100-200$ g/kg higher to produce a similar effect on $k_2$, depending on the exact $\gamma$ (Fig.~\ref{fig:dk2_mod}). Previous studies have argued in favor of a heavy convective ocean that holds $S\sim200$ g/kg in salts to obtain a $1\sigma$ agreement with $k_2$ observations. However, water-rock interactions at the bottom of Titan's ocean are expected to produce a limited $S\lesssim 10$ g/kg of salts \citep{leitner2019modeling}.

Instead of water-rock interactions, the heavy convective ocean relies on a special event to attain its relatively high salt concentration. Ammonium sulfate ((NH$_4$)$_2$SO$_4$) could form in the ocean from reactions between water-ammonia and brine leaching upwards from a core experiencing hydration \citep{fortes2007ammonium}, contributing to $S\sim 200$  g/kg of dissolved solute.  Unfortunately, this scenario leads to predicted surficial expressions on Titan that failed to be observed during the Cassini mission \citep{leitner2019modeling}. Alternatively, magnesium sulfates can be incorporated into a heavy ocean that is thermodynamically consistent \citep{vance2013thermodynamic} via a late-delivery of salt-rich carbonaceous chondrites \citep{hogenboom1995magnesium}. However, it is not clear whether this delivery mechanism can provide the required salt concentration.

Water-rock interactions on Titan's ocean floor \citep{leitner2019modeling} produce enough salts ($S\sim10$ g/kg) to enable resonant stratification. At $S\sim10$~g/kg salinity, the predicted $k_2$ is in a $2\sigma$ agreement with the Cassini observation when $\gamma=10^{-9}$~s$^{-1}$ and within $1\sigma$  agreement when $\gamma=3.3\times10^{-10}$~s$^{-1}$ (Fig.~\ref{fig:dk2_mod}). Both $\gamma$ values are realistic and prevent nonlinear wave breaking. This bulk concentration of salts ($S\sim10$ g/kg) is compatible with the average salinity of Earth's oceans and the salinity inferred for Enceladus' ocean from direct sampling of E-ring particles provided by Enceladus' plumes \citep{postberg2009sodium}. This aspect favors resonant stratification over a heavy ocean because water-rock interactions are better understood than the special event required to explain a high concentration of salts, namely early hydration during the interior differentiation of ices and silicates \citep{fortes2007ammonium} or a late delivery of carbonaceous chondrites with a special composition \citep{hogenboom1995magnesium}.





\subsection{Ocean stability to overturning convection}

 The heat flux at the ocean floor provided by interior heating threatens the stability of a weakly stratified ocean. A thermal gradient can counter the stratification produced by a chemical gradient and lead to unstable overturning convection. Convection introduces mixing that can further erase the chemical gradient over time when no salinity forcing is introduced. From a simple balance of the density profile including thermal effects and a salinity gradient (i.e., the Ledoux instability criterion), we require
 \begin{equation}
     \Delta T\lesssim \frac{H N^2}{\alpha g}\sim 10 \textnormal{ K}
     \label{eq:ledoux}
\end{equation}
 across \deleted{the same } an ocean thickness \added{$H\sim 300$ km} to preserve the stability of the mild stratification $N^2\sim1\times10^{-8}$~s$^{-2}$ discussed before, \added{where $\alpha\sim 2\times10^{-4}$ K$^{-1}$ is the thermal expansivity and $g\approx 135$ cm s$^{-2}$ is the surface gravity}. This temperature gradient is equivalent to a heat transfer ${\lesssim}2\times10^{9}$~W across the ocean by thermal conduction, two orders of magnitude lower than the heating rate expected from radiogenic heating (Appendix~\ref{app:energy}). 
 
 One might then presume that the heat flux from radiogenic heating should destroy any mild stable stratification. However, this need not necessarily be the case.  For example, plumes from hydrothermal vents or volcanoes on the surface of the silicate core may pierce through the stratified ocean in a Rayleigh-Taylor instability \citep{collins2007enceladus} and allow the radiogenic heat from the solid interior to escape outward without triggering overturning convection at the scale of the entire ocean. In this hypothetical scenario, heat passes across the ocean in small lengthscale plumes that do not disturb the large lengthscale structure of stratification required by resonant stratification.

Double-diffusive convection \citep{radko2013double} constitutes another hypothetical scenario that may permit maintenance of the radiogenic heat flux without erasing the compositional gradient. This regime is typically observed when the temperature profile is steeper than the convective temperature profile and less steep than the Ledoux instability criterion \citep[e.g.]{leconte2012new}. The convective temperature profile prescribes a $\Delta T\sim \alpha HgT/c_p\sim5.2$ K \citep{turcotte2002geodynamics} over an ocean \replaced{thickness $H\sim 300$ km}{with the same properties used in equation (\ref{eq:ledoux})}, where $c_p\sim 4$ J~g$^{-1}$~K$^{-1}$ is the heat capacity\deleted{and $\alpha\sim 2\times10^{-4}$ K$^{-1}$ is the thermal expansivity}. Assuming thermal equilibrium between conduction through the ocean thickness and the radiogenic heat flux (Appendix~\ref{app:energy}), we require
\begin{equation}
    N^2 \gtrsim \frac{\alpha g \dot{E}_{int}}{4\pi \rho c_p R^2_p \kappa} \sim 1.6\times10^{-6}\left(\frac{1.5\times10^{-3}\textnormal{ cm}^2\textnormal{s}^{-1}}{\kappa}\right) \textnormal{ s}^{-2}
    \label{eq:kappa}
\end{equation}
to maintain Ledoux stability (equation~(\ref{eq:ledoux})) of the ocean salinity gradient, where $\kappa$ is the thermal diffusivity of the ocean. In regimes typical of gas giant planets, numerical simulations suggest that double-diffusive convection can increase the efficiency of heat transport by up to a factor of ${\sim}50$ compared to heat conduction \citep{rosenblum2011turbulent, mirouh2012new}. This enhancement can be thought of as being roughly equivalent to an increase in $\kappa$ that allows a salinity gradient with $N^2\sim10^{-8}$~s$^{-2}$ to remain Ledoux stable (equation (\ref{eq:kappa})). Numerical simulations confirm a strong enhancement of heat transport by double-diffusive convection in the regime of icy satellites \citep{wong2022layering}, where, contrary to the gas giant planets, the kinematic viscosity $\nu$ is typically larger than the thermal diffusivity $\kappa$. Double-diffusive convection is typically accompanied by the evolution of the compositional gradient \citep{radko2013double,wong2022layering}, thus further studies are required to better understand the timescales imposed on the evolution of salinity profiles in icy satellites.

\subsection{Predictions and future tests}

Water-rock interactions can occur at the ocean bottom of other icy satellites, thus an enhancement of $k_2$ by resonant stratification is also possible on Ganymede and Europa. NASA's Europa Clipper mission \citep{phillips2014europa,howell2020nasa} will measure Europa's $k_2$ with an expected accuracy of $2\%$ \citep{mazarico2023europa}, providing us with a new opportunity to study the interior of an icy satellite. If resonant stratification is a common mode of operation for icy satellites, we should also observe an important enhancement in Europa's $k_2$ given that the ice shell elasticity only provides small resistance to vertical tidal displacements. ESA's JUICE mission \citep{grasset2013jupiter} will measure Ganymede's $k_2$ with even greater precision due to the orbital design of the mission and the improved K-band antenna onboard the spacecraft. Ganymede's tidal response will be measured at the excitation frequency of the various moon-to-moon tides present in the Galilean moon system \citep{DeMar-etal:2022}, in addition to the conventional eccentricity tide raised by Jupiter. This tidal response spectrum will provide a unique opportunity to measure the potential stratification of an ocean regardless of whether resonant stratification is in place or not, in addition to providing further constraints to ocean thickness from sampling high-frequency moon-to-moon tides. One quantity to look for in the tidal response spectrum is the $g$-mode spacing, which is both sensitive to the degree of stratification and the thickness of the stratified cavity (Appendix~\ref{app:modes}). 

\section{Conclusions}

We calculated Titan's tidal response to eccentricity tides using a new theoretical framework that includes the dynamical effects of tidally excited waves trapped in the ocean. Our results present a new interpretation of Cassini's observation of Titan's Love number $k_2=0.616\pm0.067$, which is $3-\sigma$ away from the predicted $k_2$ in an ocean of pure water resting on top a rigid ocean floor. If Titan's ocean is stably stratified, its measured tidal response can be fully reproduced using plausible dissipation factors ($\gamma$) without requiring a salinity greater than those of Earth or Enceladus's oceans (Fig.~\ref{fig:dk2_mod}). This enhanced response requires the ocean to be set in resonance with the period of the current tidal excitation, namely Titan's orbital period. In one possible scenario, this resonance is encountered as the ocean progressively freezes and develops a deep salty layer (Fig.~\ref{fig:gmodes}); this situation yields a long-term stable thermal equilibrium with conduction across the ice shell. 
 
Studies on the extent to which stably stratified oceans can be maintained against convective mixing would form a valuable theoretical addition to the current work. Processes similar to those hypothesized to be operating at Titan could be in play at Ganymede or Europa, and may be tested with future spacecraft missions Europa Clipper and JUICE. The seismometer expected on the Dragonfly mission \citep{Barne-etal:2021}, or a future Titan orbiter \citep[e.g.]{Sotin-etal:2017}, might similarly be able to look for evidence of a resonantly-excited ocean.

\acknowledgments
We thank David Stevenson for insightful early discussions leading to this manuscript. We are thankful for the constructive comments of Mikael Beuthe and one anonymous referee.

%

\vspace{5mm}


\software{Mathematica \citep{wolfram1999mathematica}, Matplotlib \citep{hunter2007matplotlib}
          }



\appendix

\section{The spectrum of internal gravity waves in a stratified ocean \label{app:modes}}
In the absence of rotation, the spectrum of normal modes associated with internal gravity waves (e.g., g-modes) follows \citep{unno1979nonradial}
\begin{equation}
    _n\omega_\ell \simeq \frac{\sqrt{\ell(\ell+1)}}{n\pi}\int_0^{R}\frac{N}{r}dr\textnormal{,}
    \label{eq:gmodes}
\end{equation}
where $n$ is the overtone radial order, $\ell$ is degree, and $N$ the Brunt-Vaisala frequency. For Titan, we can estimate the $N$ that produces a resonance with eccentricity tides of tidal frequency $\omega=\omega_s$ using our simplified linear model of stratification (i.e., constant $N$ across a stratified cavity of thickness $H$),
\begin{equation}
    N = -0.424\left(\frac{n\pi}{\ln(1-H/R)\sqrt{\ell(\ell+1)}}\right) \omega_s\textnormal{.}
    \label{eq:gmodes_N}
\end{equation}
We obtain the numerical factor in front of the previous expression from a fit of equation~(\ref{eq:gmodes}) to individual resonances identified in our numerical simulations (Fig.~\ref{fig:gmodes}). This factor represents the shift caused on the g-mode frequency due to the effects of Titan's rotation with spin rate $\Omega=\omega_s$. 

The g-mode frequency spacing constitutes a diagnostic quantity typically used to characterize stratified cavities inside planets and stars \citep[e.g.]{aerts2010asteroseismology,mankovich2021diffuse}. Following our simple model of constant $N$ across a stratified cavity of thickness $H$, we obtain the mode spacing
\begin{equation}
    \Delta P_g \sim -0.424\left(\frac{2\pi^2}{N\ln(1-H/R)\sqrt{\ell(\ell+1)}}\right)\textnormal{.}
\end{equation}
This expression can be used to characterize the stratification of oceans in icy satellites when a multi-frequency $k_2$ is available to observation, as currently expected from JUICE measurements of moon-to-moon tides on Ganymede \citep{DeMar-etal:2022}. Future icy satellite seismology could provide an alternative observation of mode spacing from the recording of free oscillations on the satellite's surface \citep{marusiak2021exploration}, assuming that the normal modes are excited beyond the detection threshold.

\section{The tidal forcing of eccentricity tides \label{app:forcing}}

The gravitational potential $\phi^T$ experienced by an observer at $\mathbf{r}$ from a concentrated mass located at $\mathbf{r}'$, is inversely proportional to the distance between them
\begin{equation}
     \phi^T \propto \frac{1}{|\mathbf{r} - \mathbf{r}'|} = \sum_{\ell=0}^\infty \frac{r^\ell}{r'^{\ell+1}}\mathcal{P}_\ell(\cos\alpha)\textnormal{,}
\end{equation}
where $\alpha$ is the angle between the two position vectors, $\ell$ is degree, and $\mathcal{P_\ell}$ are the $m=0$ case of the associated Legendre polynomials
\begin{equation}
\mathcal{P}_\ell^m(\mu) = \frac{(-1)^m}{2^\ell \ell!}(1-\mu^2)^{m/2}\frac{d^{\ell+m}}{d\mu^{\ell+m}}(\mu^2-1)^\ell\textnormal{,}
\end{equation}
where $m$ is azimuthal order. When the concentrated mass $M$ is that of a planet orbiting at a semi-major axis $a$, the gravitational excitation assumes the form
\begin{equation}
    \phi^T = \frac{\mathcal{G}M}{a} \sum_{\ell=2}^\infty \left(\frac{r}{a}\right)^\ell \mathcal{P}_\ell(\cos\alpha )\textnormal{.}
\end{equation}
The two lowest degree harmonics, $\ell=0$ and $\ell=1$, are discarded as they do not disturb the shape of the icy satellite. 

The addition theorem allows us to express the gravitational excitation in spherical coordinates, following
\begin{equation}
    \cos\alpha = \cos\theta\cos\theta_p + \sin\varphi\sin\varphi_p \cos(\varphi - \varphi_p)\textnormal{,}
\end{equation}
where $r\theta\varphi$ are spherical polar coordinates in a corotating frame fixed to the icy satellite and the subscript $p$ denotes the position of the planet. The gravitational excitation is now
\begin{equation}
    \phi^T = \frac{\mathcal{G}M}{a} \sum_{\ell=2}^\infty \frac{4\pi}{2\ell + 1}\left(\frac{r}{a}\right)^\ell \sum_{m=-\ell}^\ell Y_\ell^m(\theta,\varphi)Y_\ell^{m*}(\theta_p,\varphi_p)\textnormal{,}
\end{equation}
where we use the conventional definition of spherical harmonics
\begin{equation}
    Y_\ell^m(\theta,\varphi) =  \left(\frac{(2\ell+1)}{4\pi}\frac{(\ell-m)!}{(\ell+m)!}\right)^{1/2}\mathcal{P}_\ell^m(\cos\theta)e^{im\varphi}\textnormal{.}
    \label{eq:sph}
\end{equation}

From fundamental identities of spherical harmonics, it can be shown that
\begin{equation}
    Y_\ell^{m*}(\theta_p,\varphi_p)=\left(\frac{(2\ell+1)}{4\pi}\frac{(\ell-m)!}{(\ell+m)!}\right)^{1/2} P_\ell^{m}(\cos\theta_p)e^{-im\varphi_p}\textnormal{.}
\end{equation}
The tidal forcing potential for $\ell m$ then becomes
\begin{equation}
    \phi^T_{\ell m} = \frac{\mathcal{G}M}{a} \left(\frac{r}{a}\right)^\ell \left(\frac{4\pi}{2\ell + 1}\frac{(\ell-m)!}{(\ell+m)!}\right)^{1/2} \mathcal{P}_\ell^{m}(\cos\theta_p)e^{-im\varphi_p} Y_\ell^m(\theta,\varphi) \textnormal{.}
    \label{forcing}
\end{equation}

In the standard case of a coplanar circular orbit, we have $\cos\theta_p=\varphi_p=0$, leading to 
\begin{equation}
    \phi^T_{\ell m} = U_{\ell m} \left(\frac{r}{R}\right)^\ell  Y_\ell^m(\theta,\varphi)e^{-i\omega t} 
    \textnormal{,}
\end{equation}
with the normalization constant
\begin{equation}
    U_{\ell m} = \left(\frac{\mathcal{G}M}{a}\right) \left(\frac{R}{a}\right)^\ell\left( \frac{4\pi (\ell-m)!}{(2\ell + 1)(\ell+m)!}\right)^{1/2} \mathcal{P}_l^{m}(0)\textnormal{.}
    \label{eq:U}
\end{equation}

The effect of eccentricity imposes a change in the semimajor axis and a libration in the position of the planet as seen from the corotating frame on the icy satellite,
\begin{equation}
    a = a_0(1 -e\cos\omega t)\textnormal{,}
\end{equation}
\begin{equation}
    \varphi_p = 2 e \sin \omega t\textnormal{.}
\end{equation}
Our strategy is now to expand equation (\ref{forcing}) to first order in $e$. The semimajor axis dependency expands as
\begin{equation}
    \frac{1}{a^{\ell+1}}\approx \frac{1}{a_0^{\ell+1}}\left(1+(\ell+1)e\cos\omega t \right)\textnormal{,}
\end{equation}
and the libration of the planet expands as
\begin{equation}
    e^{-im\varphi_p}\approx 1 - 2ime\sin\omega t \textnormal{.}
\end{equation}
Using the standard trigonometric definitions
\begin{equation}
    \sin z = \frac{e^{iz}-e^{-iz}}{2i}\textnormal{,}
\end{equation}
\begin{equation}
    \cos z = \frac{e^{iz}+e^{-iz}}{2}\textnormal{,}
\end{equation}
we obtain
\begin{equation}
    \frac{1}{a^{\ell+1}}\approx \frac{1}{a_0^{\ell+1}}\left(1+\frac{e(\ell+1)}{2}(e^{i\omega t}+e^{-i\omega t}) \right)\textnormal{,}
\end{equation}
\begin{equation}
    e^{-im\varphi_p}\approx 1 - em(e^{i\omega t}-e^{-i\omega t})\textnormal{.}
\end{equation}
The resulting tidal excitation potential in a planar orbit is
\begin{equation}
    \phi^T_{\ell m} \approx  U_{\ell m} \left(\frac{r}{R}\right)^\ell  \left(1 +  e\left(\frac{(\ell+1)}{2} -m\right) e^{i\omega t} + e\left(\frac{(\ell+1)}{2} +m        \right)e^{-i\omega t}   \right)Y_\ell^m(\theta,\varphi)  \textnormal{.}
\end{equation}

The circular tide becomes static (i.e., no time dependence) in a synchronous corotating frame where the spin of the icy world matches the orbital frequency of the planet. Eccentricity tides propagate at the diurnal frequency in both west and east directions for a given $m$. We observe a perfect superposition between the east $m>0$ tide and the west $m<0$ tides. As a result, the contributions are typically added to obtain
\begin{equation}
    \phi^T_{\ell m} \approx U_{\ell m} \left(\frac{r}{R}\right)^\ell  Y_\ell^m(\theta,\varphi) \left(1 +  e\left(\frac{(\ell+1)}{2} -m\right) e^{i\omega t} + e\left(\frac{(\ell+1)}{2} +m        \right)e^{-i\omega t}   \right) \textnormal{.}
\end{equation}

We make use of the fact that the direction of tides can be flipped by either changing the sign in $\varphi$ or the tidal frequency $\omega_s$, reducing the eccentricity tidal potential to
\begin{equation}
    \phi^{e}_{\ell m} \approx e\left(\ell+1 +2m        \right)U_{\ell m} \left(\frac{r}{R}\right)^\ell  Y_\ell^m(\theta,\varphi) e^{-i\omega t}  \textnormal{.}
    \label{eq:forcing}
\end{equation}
In this paper, all quantities mimic the time dependence of the tidal forcing. East tides correspond to $m>0$ and west tides to $m<0$. East tides propagate in the direction of rotation, whereas west tides are counter-rotation. Notice that the amplitude of eccentricity tides is $7e$ fold compared to the amplitude of the static tides in the case $\ell=m=2$.  

\section{Projection of dynamical tides onto vectorial spherical harmonics\label{app:proj}}

We project our equations onto vectorial spherical harmonics following the standard decomposition
\begin{equation}
    \bm{\xi} = y_1 {\bf Y} + y_2 {\bf \Psi} + y_3 {\bf \Phi}
    \textnormal{,}
    \label{eq:proj}
\end{equation}
where ${\bf Y}{\bf \Psi}{\bf \Phi}$ constitute an orthonormal base for the projection of vectorial fields in  spherical polar coordinates. VSH relate to scalar spherical harmonics (equation~(\ref{eq:sph})) as
\begin{equation}
    {\bf Y} = Y\hat{\bf r}
    \textnormal{,}
\end{equation}
\begin{equation}
    {\bf \Psi} = r\nabla Y
    \textnormal{,}
\end{equation}
\begin{equation}
    {\bf \Phi} = r\nabla\times {\bf Y}
    \textnormal{,}
\end{equation}
where spherical harmonics satisfy $r^2\nabla^2Y=-\ell(\ell+1)Y$. We further project the gravity response and pressure-gravity potentials, respectively, as
\begin{equation}
    \tilde{\phi}' = \tilde{\phi}'(r) Y\textnormal{,}
\end{equation}
\begin{equation}
    \psi = \psi(r) Y\textnormal{.}
\end{equation}

The projection of the continuity equation (equation~(\ref{eq1:continuity})) leads to
\begin{equation}
y_2^\ell = \frac{\partial_r\left( r^2 y_1^\ell\right)}{  \ell(\ell+1) r }
\textnormal{,}
\label{eq:cont}
\end{equation}
where the upper index on $y(r)$ denotes degree instead of exponent. This equation sets a requirement for the radial and spheroidal displacement fields. The toroidal displacement field is automatically continuous given that it constitutes a rotor on displacement rather than a relocation of fluid. 

The projection in equation~(\ref{eq:proj}) is ideal for solving the vorticity equation (equation~(\ref{eq:vort})) given the following set of convenient properties (see also \cite{rieutord1987linear}),
\begin{equation}
    \nabla\times\left(y {\bf Y} \right) = \frac{y}{r}{\bf \Phi}
    \textnormal{,}
\end{equation}
\begin{equation}
    \nabla\times\left( y {\bf \Psi}\right) = - \frac{\partial_r(ry)}{r}{\bf \Phi}
    \textnormal{,}
\end{equation}
\begin{equation}
    \nabla\times\left( y {\bf \Phi}\right) = \frac{\ell(\ell+1)}{r} y{\bf Y} + \frac{\partial_r(ry)}{r}{\bf \Psi}
    \textnormal{.}
\end{equation}
The curl of the tidal displacement in the vorticity equation can be directly obtained from these properties,
\begin{equation}
    \nabla\times \bm{\xi} = \frac{\ell(\ell+1)}{r} y_3 {\bf Y} + \frac{\partial_r(ry_3)}{r}{\bf \Psi} + \left( \frac{y_1-\partial_r(ry_2)}{r}\right){\bf \Phi}\textnormal{.}
\end{equation}
The stratification term in the vorticity equation can also be directly obtained,
\begin{equation}
    \nabla\times\left( \left(\bm{N}^2\cdot \bm{\xi} \right)\hat{\bf r}\right) = \frac{N^2 y_1}{r} {\bf \Phi}\textnormal{.}
\end{equation}

The Coriolis term in the vorticity equation requires extra effort due to the coupling it imposes on spherical harmonics of different degree (i.e., terms $\sin\theta\partial_\theta Y_\ell^m$, $\cos\theta Y_\ell^m $, and combinations of both). The partial derivatives and trigonometric functions applied to the spherical harmonics indicate changes in spherical harmonic quantum numbers described in the following recursive relations \citep{lockitch1999r},
\begin{equation}
    \sin\theta\partial_\theta Y_\ell^m   = \ell q_{\ell+1}Y_{\ell+1}^m - (\ell+1)q_\ell Y_{\ell-1}^m\textnormal{,}
    \label{eq:sindY}
\end{equation}
\begin{equation}
   \cos\theta Y_\ell^m   = q_{\ell+1}Y_{\ell+1}^m+q_\ell Y_{\ell-1}^m\textnormal{,}
   \label{eq:cosY}
\end{equation}
where
\begin{equation}
    q_\ell = \left(\frac{\ell^2 - m^2}{4\ell^2 -1}\right)^{1/2}\textnormal{.}
\end{equation}
We can derive properties from the recursive coupling relations in equations (\ref{eq:sindY}) and (\ref{eq:cosY}) that permit the projection of the Coriolis term onto radial functions,
\begin{equation}
    \bm{\Omega}\times\left(y_1 {\bf Y}\right) = -\left(\frac{im\Omega}{\ell(\ell+1)}\right)y_1{\bf \Psi} + \frac{\Omega}{\ell(\ell+1)}\left((\ell+1)q_\ell y_1^{\ell-1}- \ell q_{\ell+1}y_1^{\ell+1}\right){\bf \Phi}
    \textnormal{,}
\end{equation}
\begin{equation}
    \bm{\Omega}\times\left(y_2 {\bf \Psi}\right) = - im\Omega y_2 {\bm Y} - \frac{im\Omega}{\ell(\ell+1)}y_2{\bf \Psi} - \frac{\Omega}{\ell(\ell+1)}\left((\ell^2 - 1)q_\ell y_2^{\ell-1} + \ell(\ell+2) q_{\ell+1}y_2^{\ell+1}\right){\bf \Phi}
    \textnormal{,}
\end{equation}
\begin{eqnarray}
    \bm{\Omega}\times\left(y_3 {\bf \Phi}\right) = \Omega \left( (\ell-1)q_\ell y_3^{\ell-1} - (\ell+2)q_{\ell+1} y_3^{\ell+1}\right){\bf Y} - \frac{im\Omega}{\ell(\ell+1)}y_3{\bf \Phi}\nonumber\\ + \frac{\Omega}{\ell(\ell+1)} \left((\ell^2-1)q_\ell y_3^{\ell-1}+ \ell(\ell+2)q_{\ell+1}y_3^{\ell+1}\right){\bf \Psi}
    \textnormal{.}
\end{eqnarray}
The vorticity of the Coriolis term results in a rather long expression. For brevity, here we only show the ${\bf Y}{\bf \Phi}$ components that result from combining the vorticity and continuity equations, respectively,
\begin{equation}
      \frac{q_\ell}{\ell^2} \partial_r\left(\frac{y_1^{\ell-1}}{r^{\ell-2}}\right)r^{\ell-1} + \frac{q_{\ell+1}}{(\ell+1)^2}\frac{\partial_r\left(r^{\ell+3}y_1^{\ell+1}\right)}{r^{\ell+2}}  + \left(\frac{im}{\ell(\ell+1)} - \frac{i\tilde{\omega}}{2 \Omega}  \right)y_3 = 0 
      \textnormal{,}
      \label{eq:Ycoupled}
\end{equation}
\begin{equation}
    \frac{iN^2}{2\Omega\omega}y_1 + \left(\frac{i\tilde{\omega}}{2\Omega} - \frac{im}{\ell(\ell+1)} \right)\left( \frac{\partial_{r,r}(r^2 y_1)}{\ell(\ell+1)}- y_1 \right) + \frac{(\ell-1)q_\ell}{\ell} r^\ell\partial_r\left(\frac{y_3^{\ell-1}}{r^{\ell-1}}\right) + \frac{(\ell+2)q_{\ell+1}}{(\ell+1)}\frac{\partial_r(r^{\ell+2}y_3^{\ell+1})}{r^{\ell+1}}=0
    \textnormal{.}
\end{equation}
The first equation tells you that in the absence of rotation, the toroidal displacement is zero. We can complete the set of equations required to solve the problem by combining the momentum (equation~(\ref{eq:momentum_gmode2})) and continuity equations to produce the following ${\bf Y}$ component
\begin{equation}
\frac{\partial_r\psi}{2\omega\Omega} + \left(\frac{\omega\tilde{\omega}-N^2}{2\omega\Omega}\right)y_1 + \frac{m}{\ell(\ell+1)}\frac{\partial_r(r^2y_1)}{r}+i(\ell-1)q_\ell y_3^{\ell-1} - i(\ell+2)q_{\ell+1}y_3^{\ell+1} = 0\textnormal{.}
\end{equation}

The projection of the ocean heating rate (equation (\ref{eq:heating})) onto spherical harmonics follows a much simpler path. Using the base in equation (\ref{eq:proj}), we obtain
\begin{equation}
    \dot{E} =  \rho\gamma\omega^2 \int_{R_H}^R \left(y_1^2 +\ell(\ell+1)(y_2^2+y_3^2) \right)r^2dr\textnormal{,}
\end{equation}
where $y_1$, $y_2$, and $y_3$ are radial functions, the superscript denotes exponent, and we have used the definitions of orthonormality of vectorial spherical harmonics
\begin{equation}
    \int_{\hat{\Omega}} {\bf Y}\cdot {\bf Y} d\hat{\Omega}= 1\textnormal{,}
\end{equation}
\begin{equation}
    \int_{\hat{\Omega}} {\bf \Psi}\cdot{\bf \Psi} d\hat{\Omega}= \int_{\hat{\Omega}} {\bf \Phi} \cdot{\bf \Phi} d\hat{\Omega}= \ell(\ell+1)\textnormal{,}
\end{equation}
\begin{equation}
     \int_{\hat{\Omega}} {\bf Y}\cdot {\bf \Psi} d\hat{\Omega}=  \int_{\hat{\Omega}} {\bf Y}\cdot {\bf \Phi} d\hat{\Omega}=  \int_{\hat{\Omega}} {\bf \Psi}\cdot {\bf \Phi} d\hat{\Omega}= 0\textnormal{,}
\end{equation}
where $\hat{\Omega}$ is the solid angle and $d\hat{\Omega} = \sin\theta d\varphi d\theta$.

\section{Balance between radiogenic heating and ice shell conduction \label{app:energy}}

In the absence of ocean resonances, Titan's interior heating rate should be balanced by heat transfer across the ice shell. If we consider that Titan's ice shell operates at a temperature low enough that the most likely heat transfer mechanism at work is conduction, the heat transferred across a thin, spherically symmetric ice shell can be approximated by 
\begin{equation}
    \dot{E} \sim \frac{4\pi R^2 k \Delta T}{d}\textnormal{,}
\end{equation}
where $R$ is the mean radius of the ice shell, $k$ is the thermal conductivity of ice, $\Delta T$ is the temperature difference across the ice shell, and $d$ is the ice shell thickness. 

The main source of Titan's internal heat is radiogenic heating. The bulk composition of Titan splits into rocks and ices roughly at a mass ratio 1:1 \citep{stevenson1992interior}. We can estimate the radiogenic heating rate from terrestrial rock samples if we assume a homogeneous distribution of radiogenic elements along heliocentric distance. The abundances of Th-U-K isotopes and their radiogenic decays prescribe a heating rate $H\approx 7.4\times10^{-12}$ W/kg on Earth's mantle \citep{turcotte2002geodynamics}. The resulting radiogenic heat production is $\dot{E}_{int} \sim 0.5Hm_s \sim 5\times10^{11}$ W.

Titan's core composition most likely departs from the composition of Earth's mantle due to the presence of undifferentiated iron, in which case the previous estimate is an upper bound. 
An alternative to the previous estimate comes from using the radiogenic heat production of CV chondrites $H\approx 4.5\times10^{-12}$ W/kg \citep{grasset1996cooling,spohn2003oceans}, which results in the smaller $\dot{E}_{int}\sim3\times10^{11}$ W. 

The balance between internal radiogenic heating $\dot{E}_{int}$ and conduction across the ice shell determines the ice shell thermal equilibrium thickness. Titan's icy surface is at $T_s\sim90$ K. In a pure water ocean, we can use $k\sim 2$ W~m$^{-1}$~K$^{-1}$, Titan radius $R\approx 2575$ km, and $\Delta T\sim180$ K to obtain \citep{luan2019titan}
\begin{equation}
    d \sim \frac{4\pi R^2 k \Delta T}{\dot{E}_{int}}\sim\left(\frac{3\times10^{13}}{\dot{E}_{int}}\right)\textnormal{ km.}
    \label{eq:pureh20}
\end{equation}

Titan most likely contains a considerable amount of ammonia dissolved in its global ocean \citep{lunine1987clathrate,stevenson1992interior,grasset1996cooling}. When dissolved in water, ammonia behaves like an anti-freeze, reducing the temperature of the eutectic to $T \sim 180K$ in an ocean with 15$\%$wt. ammonia \citep{grasset1996cooling}. In this scenario, the temperature gradient across the ice shell diminishes to $\Delta T\sim 90$~K and the equilibrium ice shell thickness can be reduced in half compared to the pure water ocean (equation~(\ref{eq:pureh20})).


\bibliography{sample63}{}

\bibliographystyle{aasjournal}


\listofchanges

\end{document}